\documentclass[aps,pre,reprint,superscriptaddress,showpacs,showkeys]{revtex4-1}
\usepackage{fullpage}
\usepackage{amssymb}
\usepackage{amsmath}
\usepackage{accents}
\usepackage{subcaption}
\usepackage{ragged2e}
\DeclareCaptionJustification{justified}{\justifying}
\usepackage{graphicx}
\usepackage{color}
\usepackage{float}
\usepackage{subfig}
\usepackage{url}

\begin{document}


\title{Influence of substrate interaction and confinement on electric field induced transition in symmetric block copolymer thin films\\
	 \textcolor{red}{\textbf{(in press, Physical Review E)}}} 



\author{Arnab Mukherjee}
\email[Corresponding author: ]{muar0002@hs-karlsruhe.de (Arnab Mukherjee)}
\affiliation{Institute of Materials Processes, Karlsruhe University of Applied Sciences, Moltkestr. 30, 76133, Karlsruhe, Germany}
\affiliation{Institute of Applied Materials - Computational Materials Science, Karlsruhe Institute of Technology, Haid-und-Neu str. 7, 76131, Karlsruhe, Germany}
\author{Rajdip Mukherjee}
\affiliation{Department of Materials Science and Engineering, Indian Institute of Technology Kanpur, 208016, Kanpur, India}
\author{Kumar Ankit}
\affiliation{Institute of Applied Materials - Computational Materials Science, Karlsruhe Institute of Technology, Haid-und-Neu str. 7, 76131, Karlsruhe, Germany}
\author{Avisor Bhattacharya}
\affiliation{Institute of Materials Processes, Karlsruhe University of Applied Sciences, Moltkestr. 30, 76133, Karlsruhe, Germany}
\affiliation{Institute of Applied Materials - Computational Materials Science, Karlsruhe Institute of Technology, Haid-und-Neu str. 7, 76131, Karlsruhe, Germany}
\author{Britta Nestler}
\affiliation{Institute of Materials Processes, Karlsruhe University of Applied Sciences, Moltkestr. 30, 76133, Karlsruhe, Germany}
\affiliation{Institute of Applied Materials - Computational Materials Science, Karlsruhe Institute of Technology, Haid-und-Neu str. 7, 76131, Karlsruhe, Germany}


\date{\today}

\begin{abstract}
In the present work, we study morphologies arising due to competing substrate interaction, electric field and confinement effects on a symmetric diblock copolymer. We employ a coarse grained non-local Cahn-Hilliard phenomenological model taking into account the appropriate contributions of substrate interaction and electrostatic field. The proposed model couples the Ohta-Kawasaki functional with Maxwell equation of electrostatics, thus alleviating the need for any approximate solution used in previous studies. We calculate the phase diagram in electric field-substrate strength space for different film thicknesses. In addition to identifying the presence of parallel, perpendicular and mixed lamellae phases similar to analytical calculations, we also find a region in the phase diagram where hybrid morphologies (combination of two phases) coexist. These hybrid morphologies arise either solely due to substrate affinity and confinement or are induced due to the applied electric field. The dependence of the critical fields for transition between the various phases on substrate strength, film thickness and dielectric contrast is discussed. Some preliminary 3D results are also presented to corroborate the presence of hybrid morphologies. 
\end{abstract}


\maketitle 


\section{Introduction}\label{sec1}
Self-assembly of block copolymers has been an actively pursued field of study because of its wide technological implications \cite{bates1991polymer, bates1990block}. Depending upon the volume fraction of the components and segregation regime, block copolymers exhibit a range of periodic morphologies such as lamellae, gyroids, cylinders, spheres etc.\cite{bates1991polymer, bates1990block} As we intend to study symmetric diblock copolymers which order into lamellar structure, the rest of the paper will focus only on this particular morphology.

\begin{figure}[ht]
\includegraphics[scale=0.3]{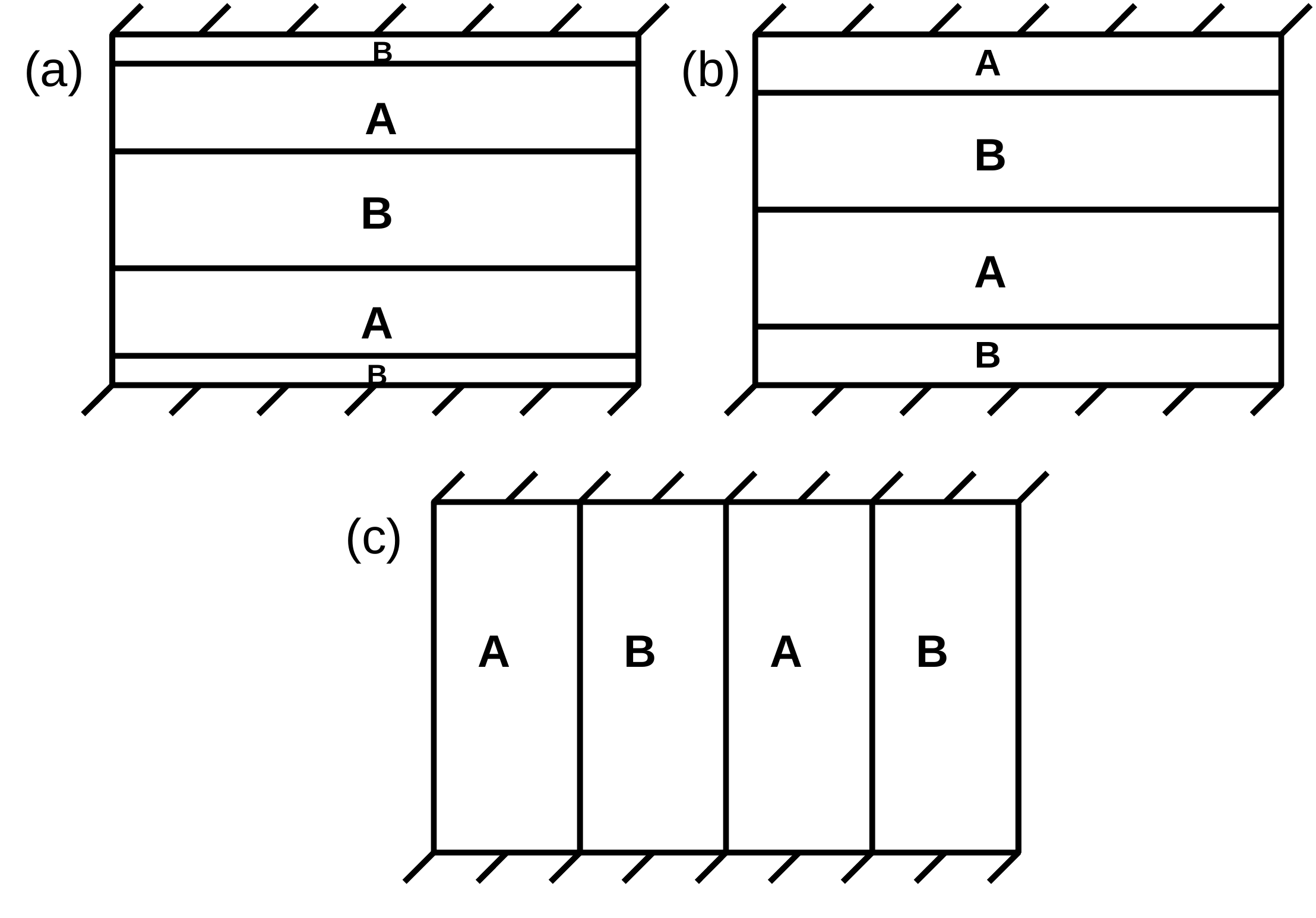}
\caption{\label{fig1} A schematic showing the various stable configurations in a symmetric diblock copolymer in presence of parallel walls/substrate (denoted by hatched lines). (a) Parallel lamellae in presence of symmetric wall interaction, (b) Parallel lamellae in presence of antisymmetric wall interaction, (c) Perpendicular lamellae arising either due to frustration or applied electric field.}
\end{figure}

In the absence of any external field, a symmetric diblock copolymer forms domains of lamellar morphology with various degrees of alignment. Practical applications, however, require complete alignment of the microphase separated domains. In general this can be achieved by application of external fields, primary among them being substrate field \cite{thurn2000ultrahigh, heier1999transfer}, shear field \cite{albalak1993microphase, balsara1994situ} and electric field \cite{amundson1991effect, thurn2002pathways, boker2002large}. Since the microphase separation generally takes place on a substrate, it is impractical to assume that the experiments are devoid of surface effects. Typically, interfacial energy difference between the two blocks in contact with a substrate (i.e $\gamma_{AS} \neq \gamma_{BS}$, where $\gamma_{AS}$ and $\gamma_{BS}$ are the interfacial energies between a monomer A or B and substrate S) can cause surface induced ordering resulting in parallel arrangement of the domains with respect to the surface \cite{fredrickson1987surface, shull1992mean, brown1994surface}. If the copolymer system is confined between two rigid substrate walls, two  different interaction cases may be considered, (a) the walls are symmetric i.e. both walls attract the same monomer and (b) antisymmetric walls i.e. both walls attract different monomers \cite{turner1992equilibrium, walton1994free, kikuchi1994microphase, brown1995ordering, wang2000monte}. As shown in Fig. \ref{fig1} (a) and (b), the system either forms integral $nL_o$ or half-integral $(n + \frac{1}{2})L_o$ number of lamellae, where $L_o$ is the equilibrium lamellar spacing, depending on whether the wall is symmetric or antisymmetric respectively \cite{turner1992equilibrium, walton1994free, kikuchi1994microphase, brown1995ordering, wang2000monte}. If the film thickness is incommensurate with the lamellar spacing, the copolymers are said to be in a frustrated state \cite{lambooy1994observed}. Frustration can be accomodated by deviation of equilibrium spacing from $L_o$ to $L$ which is typical in thicker films where the deviation can span over a large number of lamellae \cite{walton1994free, lambooy1994observed}. On the contrary, frustration in thin films is relieved by a change of configuration from parallel to perpendicular lamellae \cite{turner1992equilibrium, walton1994free, kikuchi1994microphase, brown1995ordering, wang2000monte} as shown in Fig. \ref{fig1}(c). In the present work, we restrict our study to the case of symmetric walls.

Electric field, on the other hand is gaining popularity in guiding self-assembly of block copolymers because of the ease with which it can be applied, especially in thin films \cite{xu2003interfacial, xu2004electric}. In contrast to stable parallel configuration in presence of substrate interaction, presence of electric field makes the perpendicular arrangement more stable. The reason for this can be rationalised in terms of dielectric permittivity mismatch between the two monomer components \cite{landau1984electrodynamics, pereira1999diblock, tsori2009colloquium}. In electrostatics, the electric field isolines prefer to pass through the regions of higher permittivity \cite{landau1984electrodynamics, pereira1999diblock}. Since this is readily achieved in a perpendicular state, the system minimizes its free energy by adopting perpendicular configuration.

A lot of factors can aid or deter the evolving phase morphology in presence of electric field such as the segmental interaction or segregation $\chi N$ ($\chi$ is the Flory interaction parameter which scales inversely with temperature and $N$ is the degree of polymerization), the substrate interaction strength and the confinement which can act as geometric barrier. The copolymers can be easily modulated at low segregation owing to weak segmental interaction between the two blocks \cite{tsori2002thin, matsen2006electric}. The substrate interaction can hinder the alignment process by electric field as then the applied field has to overcome the interfacial interaction \cite{xu2003interfacial}. Confinement (film thickness) on the other hand can aid or hinder the alignment process by electric field depending upon its relation with the natural lamella period $L_o$ \cite{tsori2002thin, ashok2001confined}.

The existing literature on parallel to perpendicular lamella transition can be grouped into three categories depending upon the methodology of study: analytical \cite{pereira1999diblock, ashok2001confined, tsori2002thin}, experimental \cite{xu2003interfacial, xu2004electric} and numerical \cite{kyrylyuk2003simulations, lyakhova2006kinetic, lin2005structural, matsen2006electric}. While most of the earlier theoretical work in this regard focus on strong segregation limit (SSL), \cite{pereira1999diblock, ashok2001confined, tsori2002thin} literature on weak segregation limit (WSL) \cite{tsori2002thin} is relatively sparse. A common finding in the SSL is the presence of three stable configurations, namely, parallel, perpendicular and mixed with respect to the substrate. The mixed structure contains parallel lamellae closer to the substrate and perpendicular lamellae in the middle, depending upon substrate interaction strength and electric field \cite{pereira1999diblock, tsori2002thin}. If $|\gamma_{BS} - \gamma_{AS}| < \gamma_{AB}$, the electric field causes a direct transition from parallel to perpendicular state \cite{pereira1999diblock, tsori2002thin}. However, if, $|\gamma_{BS} - \gamma_{AS}| > \gamma_{AB}$, two critical fields establish in thicker films \cite{tsori2002thin}. At lower critical field, the parallel arrangement changes to mixed morphology, while, on increasing electric field results in perpendicular lamellae \cite{tsori2002thin}. In WSL, the mixed morphology is not predicted and the transition from parallel to perpendicular is direct \cite{tsori2002thin}.

Experimental studies \cite{xu2003interfacial, xu2004electric} focusing on intermediate to strong segregation regimes have reported some observations, which do not completely converge with analytical findings. It is reported that a completely perpendicular arrangement in presence of electric field can only occur when the interfacial energy difference between the two blocks with the substrate is balanced \cite{xu2003interfacial} i.e. $\gamma_{BS} = \gamma_{AS}$ . Any mismatch results in a mixed morphology \cite{xu2003interfacial}. The discrepancy between the analytical and experimental observations  are attributed to the pathway dependence of the alignment process in the presence of an electric field \cite{xu2003interfacial}. Analytical studies on the other hand focus only on the final equilibrium configuration. In reality, however, the microdomains can get kinetically trapped in metastable states and a high activation energy may be needed to reach the stable state. 

Thus, numerical studies can provide an efficient bridge between the analytical calculation and experiments. Dynamic density functional theory (DDFT)  \cite{kyrylyuk2003simulations, kyrylyuk2006electric, lyakhova2006kinetic, xu2005electric}, self-consistent-field theory (SCFT) \cite{lin2005structural, matsen2006electric} and cell dynamics simulations (CDS) \cite{pinna2008kinetic} have previously been employed to study electric field induced alignment. 

The focus of most numerical studies have either been on mechanism of alignment of ordered domains \cite{boker2003electric, zvelindovsky2003comment, schmidt2005influence, pinna2009mechanisms, ruppel2013electric} or on order-order transition from one morphology to the other in bulk samples \cite{ly2007kinetic, ly2013kinetic, li2013electric, schmidt2010electric, pester2015electric}. The effect of substrate interaction and confinement have not been investigated thoroughly which could significantly alter the phase morphologies and critical electric field for transition. Matsen studied the stability of monolayer \cite{matsen2005stability} and multilayer \cite{matsen2006undulation} lamellae films but did not consider substrate affinity. Lyakhova et al. \cite{lyakhova2006kinetic} and Kyrylyuk et al. \cite{kyrylyuk2006electric} did consider substrate interaction, but the effect of film thickness was not considered. Moreover, the presence of any intermediate or mixed phases were not observed. Thus, the primary objective of the present work is to systematically investigate the effect of substrate interaction and confinement on the resulting morphologies. We adopt a coarse-grained Cahn-Hilliard approach to study the self-assembly of symmetric diblock copolymers under competing substrate interaction, electric field and confinement. 

Secondly, one of the serious shortcoming of most previous works involve an approximate solution of the electrostatic field. The perturbed solution of the Maxwell equation is based on the assumption of weak fractional variation of the dielectric constant \cite{amundson1993alignment, amundson1994alignment}. Though appropriate in the proximity of the order-disorder transition(ODT) temperature, the results can be significantly marred as the segregation increases. Such an assumption is relaxed in the present study by coupling the Ohta-Kawasaki functional to Maxwell equation to calculate the electrostatic field distribution.

The organization of the paper is as following. In the next section  we present our diffuse interface model, followed by a presentation of our results in section \ref{sec3}. We conclude the paper by comparing the results to that from experiments and SCFT calculations in section \ref{sec4}.

\section{Theoretical Model and Numerical Methods}\label{sec2}
\subsection{Theoretical Model}
The diffuse interface approach to model block copolymers follows the Ohta-Kawasaki free energy functional \cite{ohta1986equilibrium}, which includes a long-range interaction term in addition to short range interaction terms (bulk + interfacial energy) in the Cahn-Hilliard model \cite{cahn1958free} to account for chain interactions. The bulk part of the free energy functional of a diblock copolymer can be written as \cite{bahiana1990cell, liu1989dynamics, chakrabarti1989microphase, choksi2009phase},

\begin{eqnarray}
\frac{F_{bulk}}{k_BT}  = & \int _{V} \left[ f(\psi) + \frac{\kappa}{2}|\nabla \psi| ^{2} \right] d\textbf{r} \\ \nonumber
 &+ B\int_{V} \int_{V} G(\textbf{r},\textbf{r}^\textbf{\ensuremath{\prime}})\psi(\textbf{r})\psi(\textbf{r}^\textbf{\ensuremath{\prime}})d\textbf{r} d\textbf{r}^\textbf{\ensuremath{\prime}}
\end{eqnarray}

where the terms in the first integral constitute the short-range interactions, while the second integral consists of the long-range interaction. $\psi(\textbf{r},t)$ is an order parameter that denotes the local concentration difference between the two components, $\psi_A - \psi_B$. $f(\psi)$ is the bulk free energy taken to be of the form $-\frac{\psi^2}{2} + \frac{\psi^4}{4}$ to account for two stable phases ($\psi = \pm 1$) below the critical temperature. $\kappa$ is the gradient energy coefficient which penalises gradients in the order parameter. $B$ is a numerical parameter that determines the extent of microphase separation and scales as $N^{-2}$, where N is the number of segments. G is the Green's function having the property $\nabla^2 G(\textbf{r},\textbf{r}^\textbf{\ensuremath{\prime}})$ = $-\delta (\textbf{r} - \textbf{r}^\textbf{\ensuremath{\prime}})$. The parameters $\kappa$ and $B$ are related to the polymer architecture through the relations \cite{ohta1986equilibrium, choksi2009phase},

\begin{eqnarray}
\kappa & \sim & \frac{l^2}{f(1-f)\chi} \\
B & \sim & \frac{1}{2f^2(1-f)^2l^2\chi N^2}
\end{eqnarray}

where $l$ is the Kuhn statistical length or the average monomer space size and $f = \frac{N_A}{N}$ is the relative molecular weight which is a measure of the length of A monomer chain compared to the whole macromolecule. The segregation $\chi N$ is determined by $\kappa$ and $B$ as \cite{choksi2009phase},
\begin{eqnarray}
\chi N & \sim & \frac{1}{\sqrt{2B\kappa}f^{3/2}(1-f)^{3/2}}
\end{eqnarray}

The free energy of the domain surface in presence of attracting walls is written as, \cite{schmidt1985model, tsori2001diblock}
\begin{eqnarray}
\frac{F_{surface}}{k_BT} & = & \int _{V} \left[ h(\textbf{r})\psi(\textbf{r}) + \frac{1}{2}g_s\psi^2(\textbf{r})\right] d\textbf{r} .
\end{eqnarray}
The above expression results from a Taylor series expansion of bare surface energy \cite{schmidt1985model}. The terms $h(\textbf{r})$ and $g_s$ have special physical interpretation. $h(\textbf{r})$ denotes the surface chemical potential difference. A positive value expresses preferential attraction of B component and vice-versa \cite{schmidt1985model, tsori2001diblock}. The term $g_s$ takes into account the deviation of Flory parameter $\chi$ at the surface \cite{tsori2001diblock}

To account for two confining walls at two ends in the current study, we rewrite the above expression in terms of a $\delta$-function as,
\begin{eqnarray}
\frac{F_{surface}}{k_BT} & = & \int _{V} \left[ h_{o}\delta(y) + h_{L}\delta(y-L)\right]\psi(\textbf{r})d\textbf{r} ,
\end{eqnarray}
where, $h_{o}$ and $h_{L}$ are the interaction strengths of the wall at $x = 0$ and $L$ respectively. In the present study the term $g_s$ is set to zero, i.e.  we neglect any deviation of interaction from the bulk. This specific choice of surface potential results in short range interaction. The electrostatic contribution to the free energy functional can be obtained as \cite{landau1984electrodynamics, amundson1993alignment},
\begin{eqnarray}
\frac{F_{electrostatic}}{k_BT} & = & - \frac{\epsilon_ov_o^3}{k_BT}\int _{V} \frac{\epsilon(\psi)}{2}|\nabla \phi| ^2 d\textbf{r}
\end{eqnarray}
where, $\epsilon_o$ is the permittivity of free space, $v_o$ is the volume occupied by one polymer chain and $\epsilon(\psi)$ is the dielectric permittivity which is taken to be phase dependent. $\phi$ is the space dependent potential due to the applied voltage. We apply a linear interpolation of the permittivity between the two phases assuming the polymer to behave as a linear dielectric material by,
\begin{eqnarray}
\epsilon(\psi) & = & \epsilon_{A} \left(\frac{1+\psi}{2}\right) + \epsilon_{B} \left(\frac{1-\psi}{2}\right)
\end{eqnarray}
The assumption of linear dielectric behaviour has been previously employed in SCFT calculations \cite{lin2005structural, matsen2006electric}. Thus the total free energy functional in units of $k_BT$ can be finally written as,
\begin{eqnarray}
F  = F_{bulk} + F_{surface} + F_{electrostatic} .
\end{eqnarray}
Substituting the derived expressions we get,
\begin{widetext}
\begin{eqnarray}
 \frac{F}{k_BT}  = \int \left[ -\frac{\psi^2}{2} + \frac{\psi^4}{4} \right] + \frac{\kappa}{2}|\nabla \psi| ^{2} + \left[ h_{o}\delta(y) + h_{L}\delta(y-L)\right]\psi(\textbf{r}) - \frac{\epsilon_ov_o^3}{k_BT}\frac{\epsilon(\psi)}{2}|\nabla \phi| ^2  d\textbf{r}  \\ \nonumber
 + B\int \int G(\textbf{r},\textbf{r}^\textbf{\ensuremath{\prime}})\psi(\textbf{r})\psi(\textbf{r}^\textbf{\ensuremath{\prime}})d\textbf{r} d\textbf{r}^\textbf{\ensuremath{\prime}} .
\end{eqnarray}
\end{widetext}
In absence of the electrostatic free energy contribution, the model reduces to the one from Brown et al. \cite{brown1995ordering}, that was used to study surface induced ordering of block copolymers. As evident, the free energy functional is dependent upon $\psi$ , $\phi$ and their spatial derivatives. The minimization of the functional, then, requires the evaluation of the variation with respect to each individual variable i.e. $\frac{\delta F}{\delta \psi}$ and $\frac{\delta F}{\delta \phi}$ \cite{garcia2004thermodynamically, matsen2006converting}.
The kinetic evolution of the conserved order parameter $\psi$ follows the dynamics of Model B framework \cite{hohenberg1977theory},
\begin{eqnarray}\label{0}
\frac{\partial \psi}{\partial t}  =  M \nabla ^{2} \mu
\end{eqnarray}
where the chemical potential $\mu$ is defined as,
\begin{eqnarray}
\mu = \frac{\delta F}{\delta \psi} .
\end{eqnarray}
Hence the kinetic equation can be expressed as,
\begin{widetext}
\begin{eqnarray}\label{1}
\frac{\partial \psi}{\partial t}  =   M \nabla ^2 \left[ \left(-\psi + \psi ^3\right) - \kappa \nabla ^2\psi  - \frac{\epsilon_ov_o^3}{k_BT}\frac{\epsilon ^\prime (\psi)}{2}|\nabla \phi| ^2 \right] - B\psi .
\end{eqnarray}
\end{widetext}
Additionally, the variation of the functional with respect to $\phi$, assuming the electric field to be in local equilibrium, leads to \cite{garcia2004thermodynamically},
\begin{eqnarray}\label{2}
\frac{\delta F}{\delta \phi} = \nabla \cdot \left[ \epsilon_o\epsilon(\psi)\nabla\phi \right] = 0
\end{eqnarray}
The above expression is nothing but the Maxwell equation which provides the spatial distribution of $\phi$.

We remark that thermal fluctuations can be accounted by adding a noise term $\sqrt{s}\eta$ in Eq. \ref{0}, where, $s$ is the strength of the noise, the reciprocal of which is roughly equal to the quench depth and $\eta$ is noise distribution following the fluctuation-dissipation theorem $\langle \eta(\textbf{r},t)\eta(\textbf{r}^\textbf{\ensuremath{\prime}},\textbf{t}^\textbf{\ensuremath{\prime}})\rangle = -\nabla^2 \delta(\textbf{r} -\textbf{r}^\textbf{\ensuremath{\prime}} )\delta(t - t^\ensuremath{\prime}) $. It is worth mentioning that thermal fluctuations are important as far as the order-disorder transitions in weak segregation regime are concerned. However, the equilibrium morphology which is the focus of the present work is not influenced upon incorporation of stochastic noise. The present claim is corroborated by Ref. \cite{brown1995ordering}, where the effect of confining surfaces was studied and the final morphology was found to be independent of noise, both in WSL as well as SSL. Therefore, we assert that the neglect of stochastic noise in the present work is reasonably well justified.

\subsection{Numerical methods and parameters}
\begin{figure}[ht]
\includegraphics[scale=0.3]{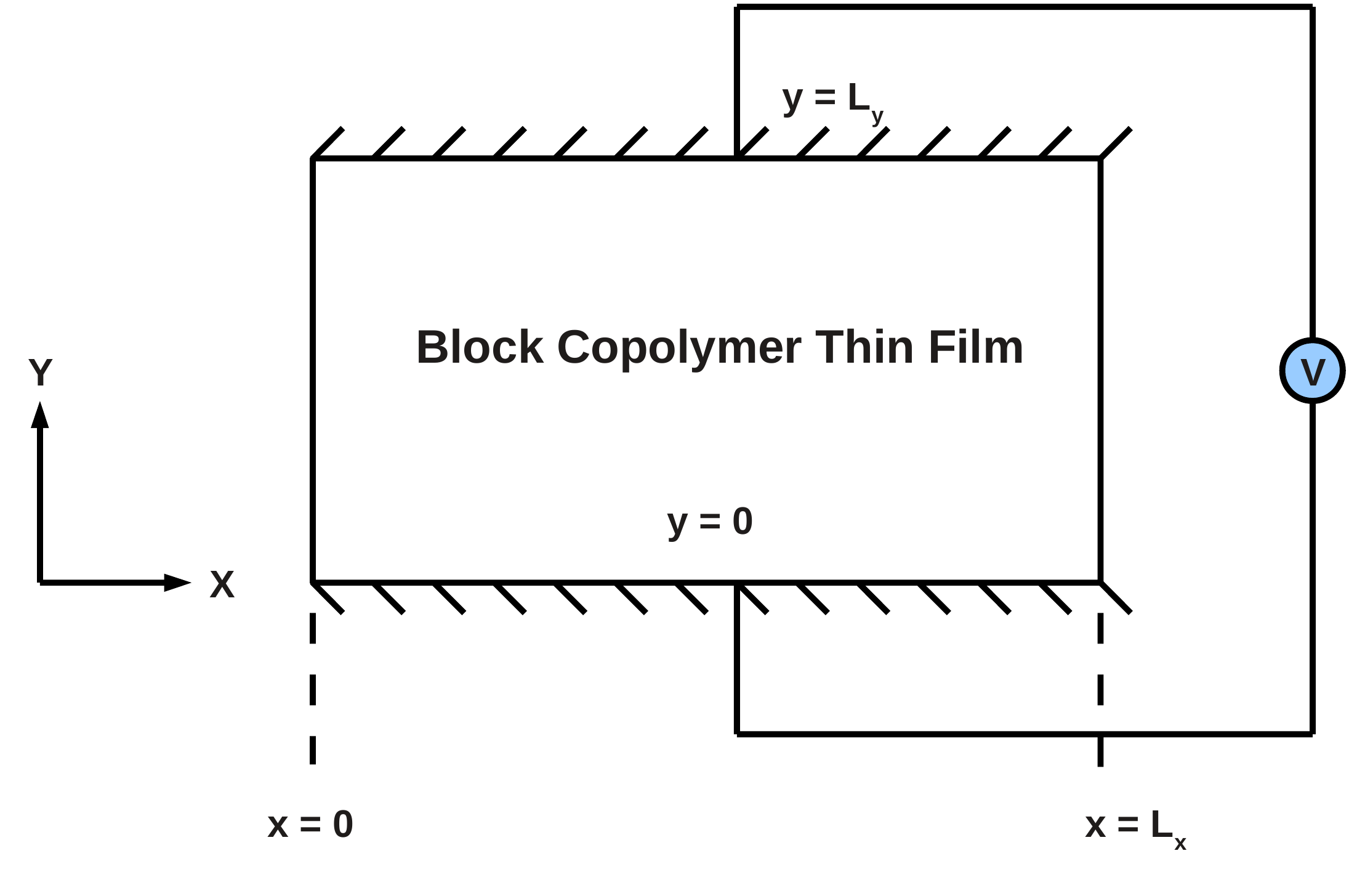}
\caption{\label{fig2} A schematic showing the simulation set up used in the present study. A diblock copolymer thin film is confined within two rigid substrates (top and bottom) which has a preferential attraction towards one of the components. Electrostatic field is generated by applying a constant voltage across the film thickness.} 
\end{figure}
We non-dimensionalize all the quantities selecting characteristic energy scale $F^\prime$, length scale $L^\prime$ and time $t^\prime$, The free energies are rescaled using $F^\prime = k_BT$, $L^\prime$ by the lattice cell size and the time by $t^\prime = \frac{{L^\prime}^2}{M^\prime}$. The terms $\kappa$ and $B$ is non-dimensionalized as, $\kappa = \frac{\kappa ^\prime}{{L^\prime}^2}$ and $B = B^\prime {L^\prime}^2$. The dielectric constants are rendered non-dimensional using $\epsilon_A = \frac{\epsilon_A^\prime}{\epsilon_o}$ and the electric field by $E = \frac{E^\prime}{\sqrt{\frac{k_BT}{v_o^3\epsilon_o}}}$. Using representative values of $T = 430K$ and $v_o^3 = 100 nm^3$, we obtain $E = \frac{E^\prime}{82 V/\mu m}$.

The simulation geometry chosen for the present study is shown in Fig. \ref{fig2}. The set-up consists of two rigid surfaces at $y = 0$ and $y = L_{y}$ confining the copolymer film across which a constant voltage is applied. Dirichlet boundary conditions are applied for voltage at $y = 0$ and $y = L_{y}$ with $\phi|_{y=0} = +\frac{V}{2}$ and $\phi|_{y=L_{y}} = -\frac{V}{2}$ while Neumann boundary condition is applied at $x=0$ and $x=L_{x}$. Therefore, electric field is aligned along y-direction. The confining substrates, also attract one of the copolymers which is controlled by the numerical parameter $h$ as mentioned before. The appropriate boundary condition to account for attracting substrates translate into, \cite{cahn1977critical, puri1992surface}
\begin{eqnarray}\label{3}
\frac{\partial \psi}{\partial y}\bigg|_{y=0}  =  +\frac{h_{o}}{\kappa} \nonumber \\
\frac{\partial \psi}{\partial y}\bigg|_{y=L_{y}} = -\frac{h_{L}}{\kappa} .
\end{eqnarray}
Additionally, no mass transport is allowed through the rigid surface by applying a no-flux boundary condition at the surfaces, \cite{puri1992surface, brown1994surface}
\begin{eqnarray}
\frac{\partial \mu}{\partial y}\bigg|_{y=0}  =  \frac{\partial \mu}{\partial y}\bigg|_{y=L_{y}}  =  0 .
\end{eqnarray}
Periodic boundary condition is applied for $\psi$ in x-direction. We solve Eq. (\ref{1}) using an explicit finite difference method where the spatial derivatives are discretized using central difference which is second order accurate in space and temporal discretization using first order Euler technique. The laplace equation in Eq. (\ref{2})  is solved iteratively using Successive-Over-Relaxation (SOR) method. The initial guess for $\phi$ is tailored by providing a linear initial profile in y-direction (corresponding to constant electric field, since, $E = -\nabla\phi$) to facilitate faster convergence. The various non-dimensional model parameters are selected as $\Delta x = \Delta y = 1.0$, $\Delta t = 0.02$, $\kappa = 1.0$, $B = 0.1$, $M = 1.0$, $\epsilon_{A} = 3.0$, $\epsilon_{B} = 2.0$. We remark, that the present results are not influenced by the choice of grid resolution ($\Delta{x}$ and $\Delta{y}$). To this end, we are able to replicate our numerical results (at $\Delta{x} = \Delta{y} = 1.0$) with finer grid spacing ($\Delta{x} = \Delta{y} = 0.5$). In order to scale up the time-step width $\Delta t$ which scales as $\left(\Delta{x}\right)^4$ for Cahn-Hilliard equation, we conveniently resort to a larger grid spacing. The values of the permittivity closely resembles a PS-PMMA copolymer system \cite{amundson1994alignment, ashok2001confined}, though other values have also been used in the literature \cite{lyakhova2006kinetic}. The values of $\kappa$ and $B$ correspond to a segregation of $\chi N \approx 18$.  The surface interaction strength $h$ is varied as 0.1, 0.5, 1.0 and 1.5. Since we intend to study symmetric walls i.e. both the surfaces attract the same monomer, we select $h_{o} = h_{L}$.  Moreover, we are interested to study the effect of confinement. Simulations are carried out for different box sizes in y-direction, $L_y= 64, 32, 16, 8, 4$. The natural lamellar spacing is around $10$ grid points, so that the selected film thicknesses allows us to study systems with $L_y > L_o$, $L_y \sim L_o$ and $L_y < L_o$. Another implication of the above mentioned values of substrate interaction and film thickness is that the surface induced ordering length is greater than the film thickness. In other words, this implies that in the absence of electric field, lamellae, parallel to the substrate, span across the entire film. The surface induced ordering length for the smallest substrate affinity of $h = 0.1$ is around $8L_o$. The magnitude of electric field is tuned by changing the value of applied voltage, and by normalising it with the box size $L_y$, i.e., $E = \frac{V}{L_y}$ to maintain the same electric field for different box sizes. The box size in x-direction is kept fixed as $L_x = 64$ in all the simulations. The initial microstructure is generated by assigning a computational noise between $\pm 0.005$ about the average composition ($\psi = 0$) corresponding to a disordered state. The system is then allowed to evolve in presence of electrostatic field and attracting substrates.

To gain insights during the microstructure evolution process we define two parameters, average density profile along y direction $\rho(y,t)$ \cite{brown1994surface, yan2007wetting} and degree of alignment $\beta(t)$ \cite{hori2007phase, zhang2013polymerization} as,
\begin{eqnarray}
\rho(y,t) & = & \frac{1}{L_x}\sum_{x=1}^{L_x} \psi(x,y,t) \\
\beta(t) & = & \frac{\sum_{k_x,k_y}\frac{k_{x}^2-k_y^2}{\textbf{k}^2}{S(k_x,k_y,t)}}{\sum_{k_x,k_y}S(k_x,k_y,t)}
\end{eqnarray}
where $k_x$ and $k_y$ are the fourier space wave vectors in x and y direction and $\textbf{k}^2$ = $k_x^2 + k_y^2$. $S(k_x,k_y)$ is the magnitude of intensity of the fourier power spectrum defined as $\frac{1}{L}\left\langle \psi(\textbf{k},t)\psi(\textbf{-k},t) \right\rangle$. $L$ denotes the system size $L_x \times L_y$ and the terms in the angular bracket imply the product of $\psi$ and its complex conjugate in fourier space. In cases, where the alignment is parallel to the substrates i.e parallel lamellae along the y direction form, $k_x \approx 0$, as there is no relevant periodicity along this direction. As a result, the value of degree of alignment parameter is $\beta = -1$. In the opposite case, the alignment is perpendicular to the surface, $k_y \approx 0$ and the value of degree of alignment parameter is $\beta = 1$. Thus, a value of $\beta = -1$ implies $100\%$ parallel lamellae, while, a value of $\beta = 1$ implies a $100\%$ perpendicular lamellae. A parallel lamellae arrangement in y direction is characterized by an oscillatory average density profile, whereas a flat profile about $\rho(y) = 0$ corresponds to a perpendicular lamellar arrangement.
\section{Results}\label{sec3}
\subsection{Effect of electric field and surfaces}
We first study some typical morphologies arising due to the interplay of substrate interaction, confinement and electric field followed by an evaluation of the resulting phase diagram. The result is categorized into three different regimes depending on the film thickness $L_y$.
\begin{figure*}[ht]
\includegraphics[scale=0.5]{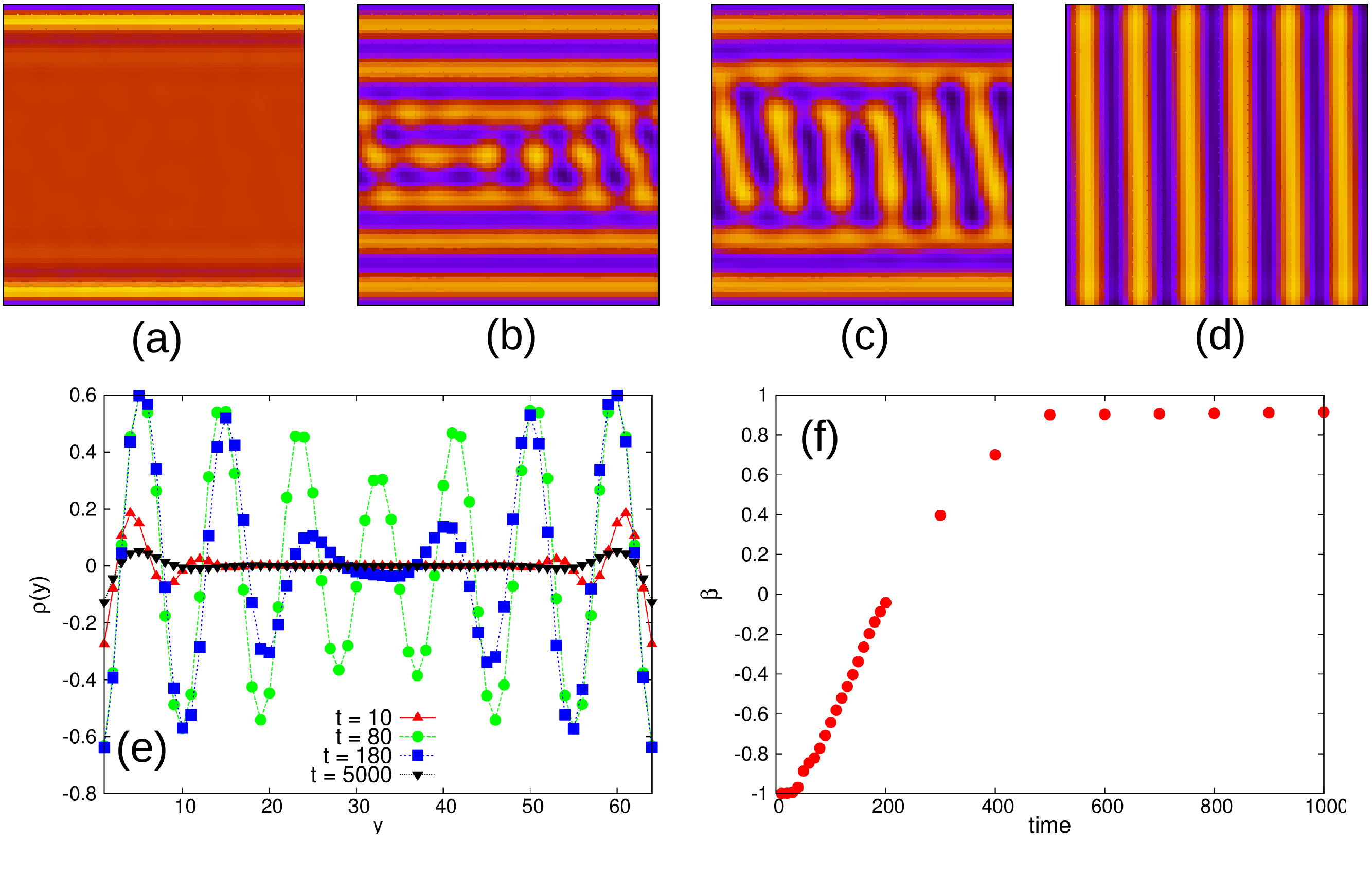}
\caption{\label{fig3} Microstructure evolution with model parameters $h = 0.1$, $E = 0.937$ and $L_y = 64(6L_o)$ at timesteps (a) t = 10, (b) t = 80, (c) t = 180 and (d) t = 5000. After the initial stages of surface induced ordering as in (a) and (b), the effect of electric field sets in resulting in a breakup of the parallel layers starting from the inner film (b) and subsequently joining in the perpendicular direction. The final stable configuration is perpendicular as shown in (d). The density profile in (e) and alignment kinetics in (f) also corroborate the fact that local break up and coalesence in the direction of applied field is the mechanism of alignment.}

\end{figure*}
\subsubsection{Thicker films with $L_y>>L_o$}
The combined effect of the substrate and electric field for model parameters $h = 0.1$, $E = 0.937$ and $L_y = 64(6L_o)$ is presented in Fig. \ref{fig3}. The phase separation initiates from the surface leading to the formation of parallel lamellae. However, at t = 80, the effect of electric field sets in, leading to undulations which ultimately break up the inner layers into smaller domains. Subsequently, the smaller domains coalesce and get aligned in the direction of the electric field. This phenomenon proceeds outwards layer by layer resulting in a perpendicular lamellar arrangement due to energetic consideration. 

The average density profile in Fig. \ref{fig3}(e) during early stages (corresponding to t = 10) is oscillatory near the surface, due to the formation of enriched and depleted layers. With time the oscillatory profile develops throughout the film thickness. At t = 180, the innermost oscillation dies out and is replaced by a flatter profile which highlights the destruction of parallel structure at the center of the film. Much later, (corresponding to t = 5000) the density profile becomes flat in the bulk of the film. However the average value shows small enrichment layers at the immediate vicinity of the surfaces even though the microstructure at the final timestep appears to be completely perpendicular. The kinetics of alignment is presented in Fig. \ref{fig3}(f). The value of $\beta$ is -1 during early stages corresponding to parallel ordering along the surface. There is a smooth temporal transition from -1 to a value closer to +0.9 which depicts the formation of perpendicular lamellae. 
\begin{figure*}[ht]
\includegraphics[scale=0.5]{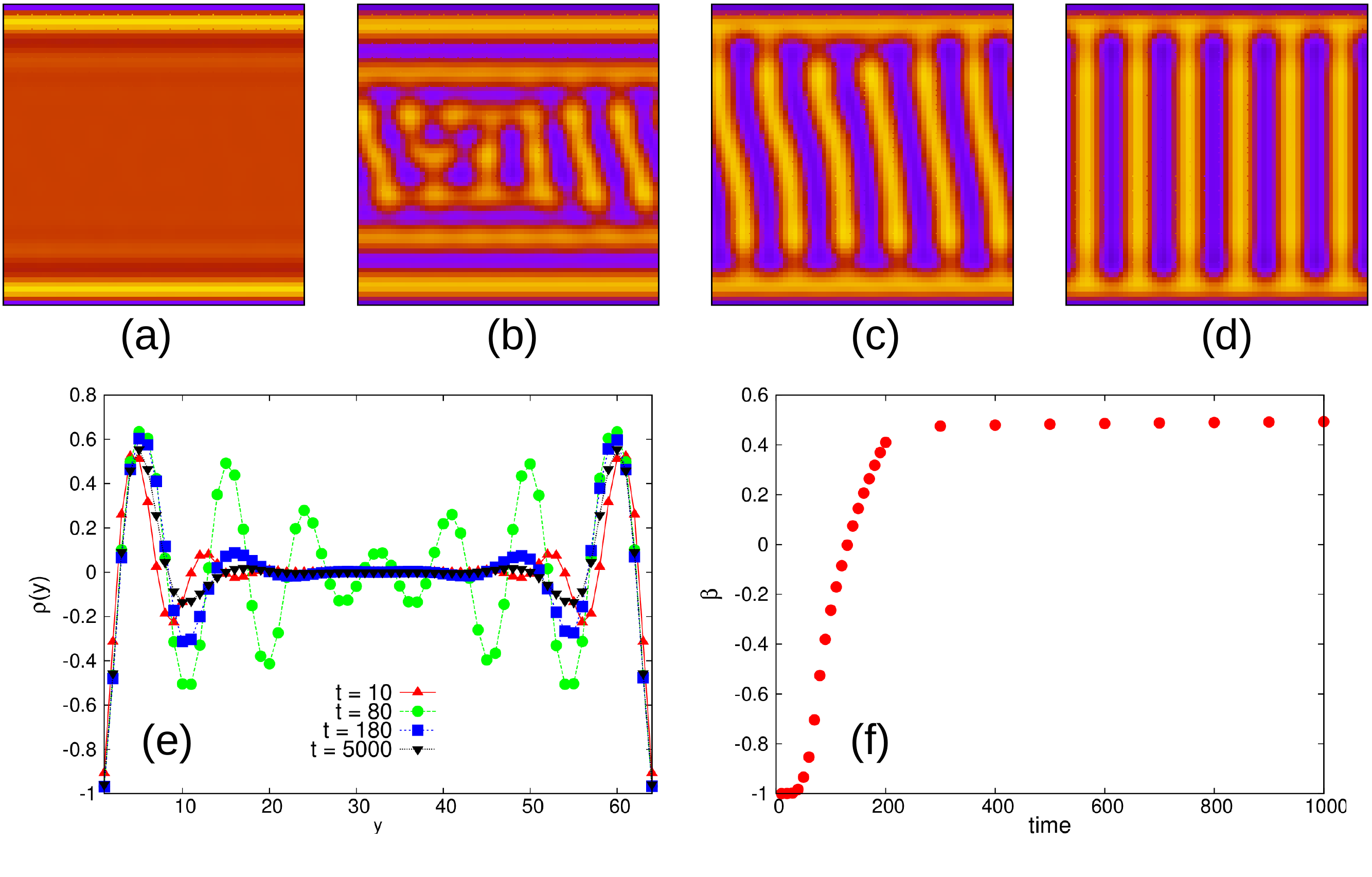}
\caption{\label{fig4}  Microstructure evolution with model parameters $h = 1.0$ and $E = 1.25$ for film thickness $L_y = 64(6L_o)$ at timesteps (a) t = 10, (b) t = 80, (c) t = 180 and (d) t = 5000. Due to the presence of higher magnitude of surface interaction strength, the lamellae near the vicinity of the surface never break, resulting in a mixed morphology mode. (e) The average density plot $\rho(y)$ consists of an oscillatory profile near the walls signifying a parallel arrangement where as the profile is flat at the middle, implying a perpendicular state. (f) The degree of alignment parameter $\beta$ captures the transition from initial parallel structure ($\beta = -1$) to the mixed state ($\beta = 0.5$).}
\end{figure*}
The influence of increasing the magnitude of substrate interaction strength and electric field for the same film thickness is presented in Fig. \ref{fig4}. The mechanism that leads to the formation of perpendicular lamellae is essentially the same as earlier, i.e local lamellae disruption and coalescence. However, as a result of greater substrate interaction, the parallel lamellae near the surfaces i.e. the wetting layer never break resulting in alternate enriched and depleted layers at the boundaries.  In literature such a morphology is termed as mixed \cite{pereira1999diblock, tsori2002thin, xu2003interfacial}. Though the surface induced ordering length is greater than the film thickness of our study, the effect of surface is predominant closer to the walls and fades as we move away from the walls. In other words this implies that the effect of surface is non uniform over the whole ordering length. As a result, when the electric field drives the domain alignment perpendicular to the surface, above a threshold interaction strength $h$, and below a threshold electric field $E$, the substrate interaction dominates near the walls resulting in a few parallel layers. Meanwhile the effect of electric field is predominant at the center (away from the wall) and is able to induce a change in configuration in this region. 

The degree of alignment achieved in the direction of electric field is around 75$\%$ in Fig. \ref{fig4}(e). Due to higher magnitude of electric field, faster kinetics is observed as can be seen by either comparing the microstructures in Fig. \ref{fig3} and Fig. \ref{fig4} or by comparing the slope of $\beta$ in Fig. \ref{fig3}(e) and Fig. \ref{fig4}(e) during the transition period. The latter has a steep transition region as compared to the smoother transition region in the former case.
\begin{figure*}[ht]
\includegraphics[scale=0.5]{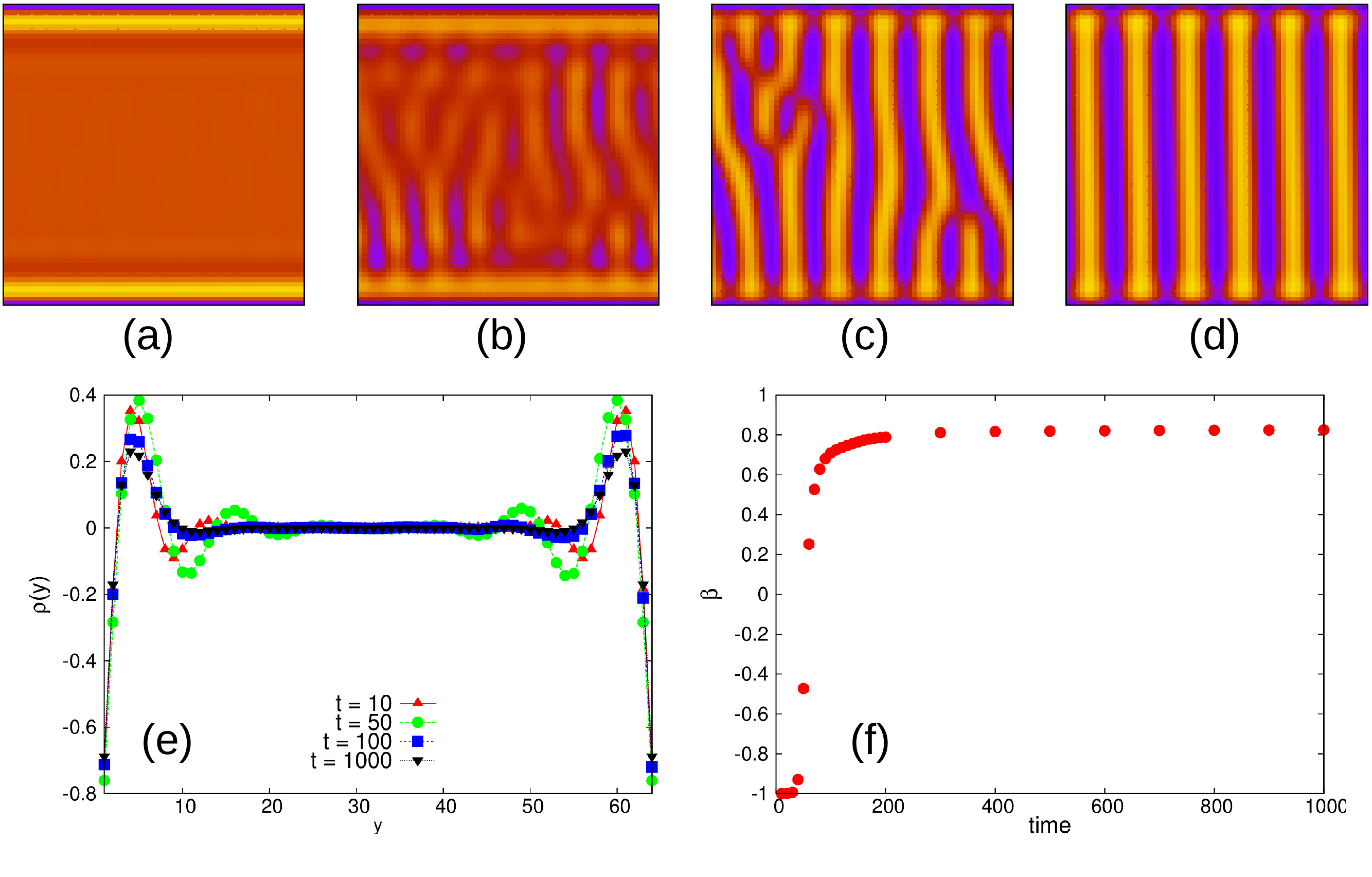}
\caption{\label{fig5}  Microstructure evolution with model parameters $h = 1.0$ and $E = 2.187$ for film thickness $L_y = 64(6L_o)$ at timesteps (a) t = 10, (b) t = 50, (c) t = 100 and (d) t = 1000. The presence of higher magnitude of electric field leads to nucleation of perpendicular layers of alternate phases at the center of the film in contrast to the mechanism of local lamellae disruption and joining at low fields.  This fact is substantiated by density profile $\rho(y)$ in (e) and alignment parameter $\beta$ (f). In the former graph, oscillatory profile characteristics of parallel configuration never develops in the middle of the film and in the latter graph the transition of $\beta$ from negative to positive values is abrupt.}
\end{figure*}
Finally we discuss the consequences of further increasing the electric field to $E = 2.187$ whilst keeping the other two parameters $h = 1.0$ and $L_y = 64$ unaltered. The results are shown in Fig. \ref{fig5}. An interesting phenomenon to observe is the mechanism of alignment by the electric field. In contrast to the previous cases, the parallel ordering never goes beyond two layers. Instead, electric field is sufficiently high to orient the composition fluctuations in the non-phase separated region, leading to the appearance of perpendicular lamellae at the middle of the film. Subsequently, the parallel layers near the walls also collapse and the rearrangement of perpendicular lamellae proceeds by defect annihilation mechanism \cite{amundson1994alignment}. The average density profile in Fig. \ref{fig5}(e) shows enrichment layers at the walls at all times but the oscillatory profile, characteristic of the parallel lamellae configuration never develops at the film center. The transition regime of alignment kinetics is abrupt as compared to the earlier cases. The value of $\beta$ saturates to a value of +0.8 which constitutes to 90$\%$ alignment in the direction of the applied field.
\subsubsection{Films with $L_y \approx L_o$}
\begin{figure*}[ht]
\includegraphics[scale=0.5]{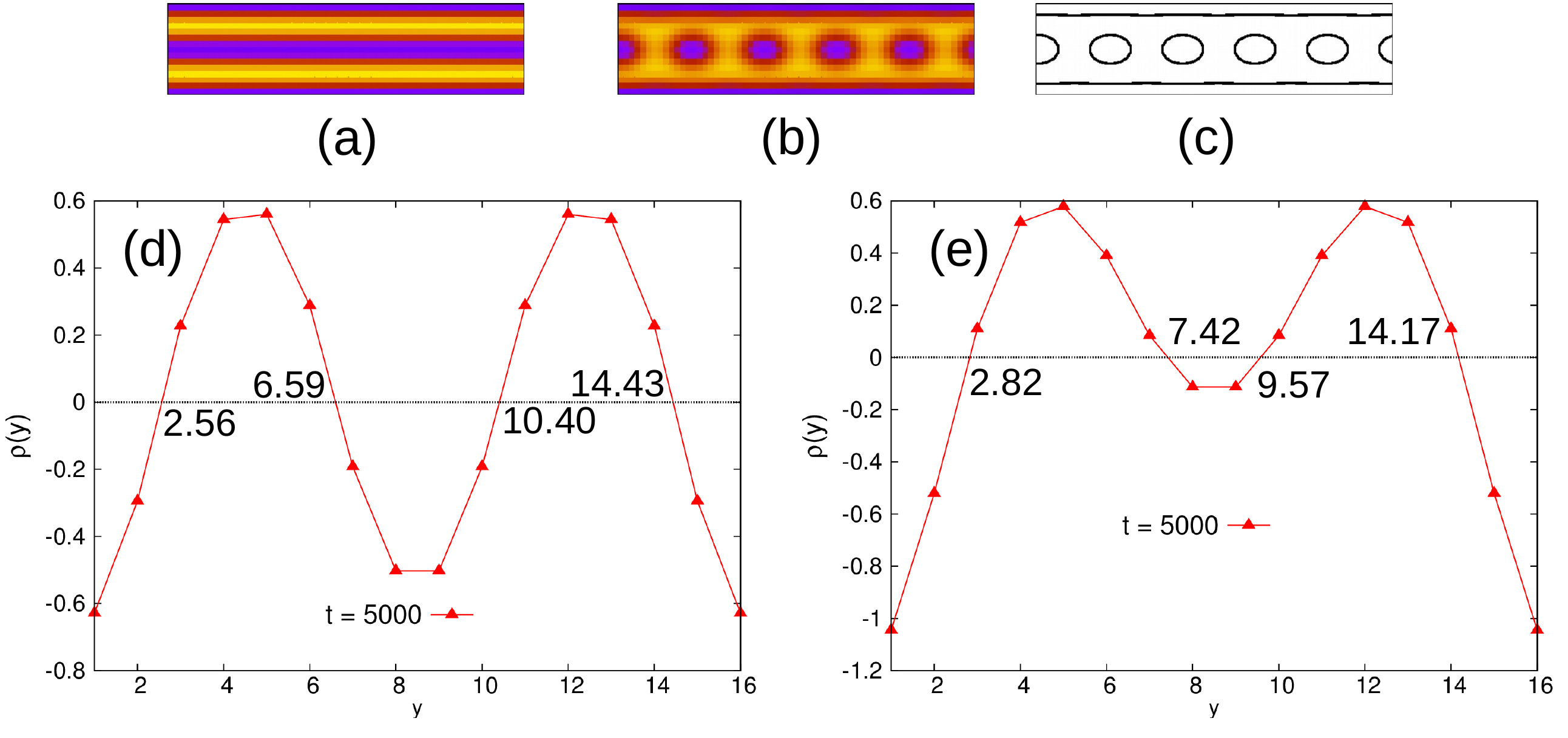}
\caption{\label{fig6} (a) microstructure corresponding to $t = 5000$ for model parameters $h = 0.1$, $E = 0.156$ for film thickness $L_y = 16(1.6L_o)$, (b) microstructure corresponding to $t = 5000$ for model parameters $h = 1.0$, $E = 0.156$ and for the same film thickness (c) contour plot corresponding to (b), (d) average density profile corresponding to (a) and (e) average density profile corresponding to (b).}
\end{figure*}
The next study focuses on configurations when the film thickness is comparable to the bulk lamellar spacing. In particular we discuss  the case of film thickness of $L_y = 16$ which is approximately equal to $1.5L_o$ . For $L_y = 16$ two cases with $h = 0.1$ (Fig. \ref{fig6}(a)) and $h = 1.0$ (Fig. \ref{fig6}(b)) are presented. The electric field is kept constant at $E = 0.156$. For low substrate interaction strengths, a parallel arrangement is found to be stable as in Fig. \ref{fig6}(a). However, with increasing substrate interaction strength we see, circular domains emerging at the center of the film as in Fig. \ref{fig6}(b). To differentiate the effect of electric field and substrate interaction responsible for this phenomenon, we consider the microstructural pattern in the absence of electric field. The resulting evolution (not shown) is similar, comprising of an inner layer of circular domains. Therefore, it can be inferred that the effect is solely driven by the substrate confinement independently of the applied electric field. The results can also be interpreted in terms of interference of composition waves. For higher $h$, the film thickness, $L_y = 16$, is close to half integral of equilibrium lamellar spacing. A destructive interference takes place at the center due to the composition waves emanating from the walls and an inner lamellae cannot be maintained. As a result circular domains start appearing in the middle. The breakup of the inner lamellae into circular domains is similar to the formation of holes as described by the mean-field theory of Shull \cite{shull1992mean}. The same phenomenon is absent at low $h$, presumably because the destructive interference at the film center is not sufficiently strong. An average density plot including the transition at $\rho(y) = 0$ for the case of lower $h$ is presented in Fig.\ref{fig6}(d). It can be verified that the innermost layer is thinner ($10.40 - 6.59 = 3.81$) than the next two adjacent layer on both sides($6.59 - 2.56 = 14.43 - 10.40 = 4.03$)  and a parallel arrangement can still be maintained with the innermost layer being in a compressed state.
\begin{figure}[hb]
\includegraphics[scale=0.5]{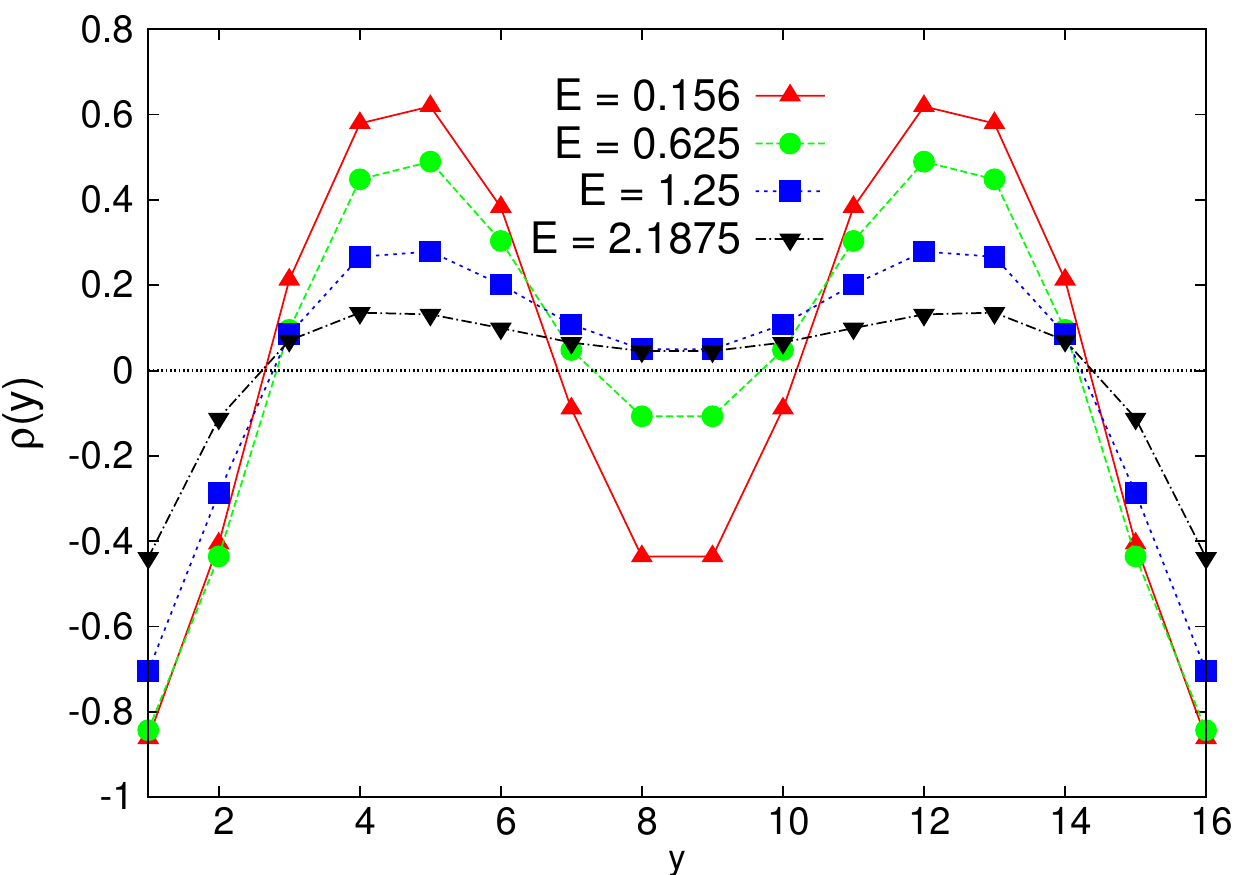}
\caption{\label{fig7} Average density profiles at different electric field strengths for $h = 0.5$. All the plots correspond to $t = 5000$.}
\end{figure}

The average density profile presented in Fig. \ref{fig6}e) shows the asymmetry in the roots of $\rho(y) = 0$. The average $\rho(y)$ value at the center is slightly B rich, clearly signifying the absence of any parallel lamellae structure. The slight asymmetry in the average density points indicates, either the evolution of a  perpendicular phase or an in plane asymmetric phase. The microstructure in Fig. \ref{fig6}b) indicates the second possibility where circular domains coexist at the center simultaneouly with the wetting layers. The presence of such hybrid structures (combination of two different phases) have previously been reported in cylinder forming systems at similar film thickness \cite{huinink2000asymmetric}. However their transition in electric field has not been reported previously. Although a 3D simulation is desirable to adequately address the issue, we can certainly make some predictions from the current 2D study. The average density profile for $h = 0.5$ at different field strengths is presented in Fig. \ref{fig7}. All the plots correspond to $t = 5000$. At low electric field strengths, $E = 0.156$, the plot is similar to Fig. \ref{fig6}d), comprising of parallel lamellae. With a slight increase in the electric field strength, the value of $\rho(y)$ at the center of the film shifts towards zero (slightly B rich). The density profile is similar to Fig. \ref{fig6}e), and denotes the appearance of a structure other than lamellae. In this case it is not the effect of substrate and confinement alone that causes this transition, but the presence of electric field does play its part. With further increase of electric field $E = 1.25$ and $2.1875$, the surface enrichment decreases. At the same time, the average value at the center of the film shifts to positive values and the profile tends to get flatter. Though it is clear that perpendicular phases now span, atleast in the middle of the film, the exact nature is very difficult to determine precisely in 2D simulation. We speculate that the perpendicular phases are either cylindrical structures (for $E = 1.25$) or perpendicular lamellae (for $E = 2.1875$). 
\subsubsection{Films with $L_y < L_o$}
\begin{figure*}[ht]
\includegraphics[scale=0.5]{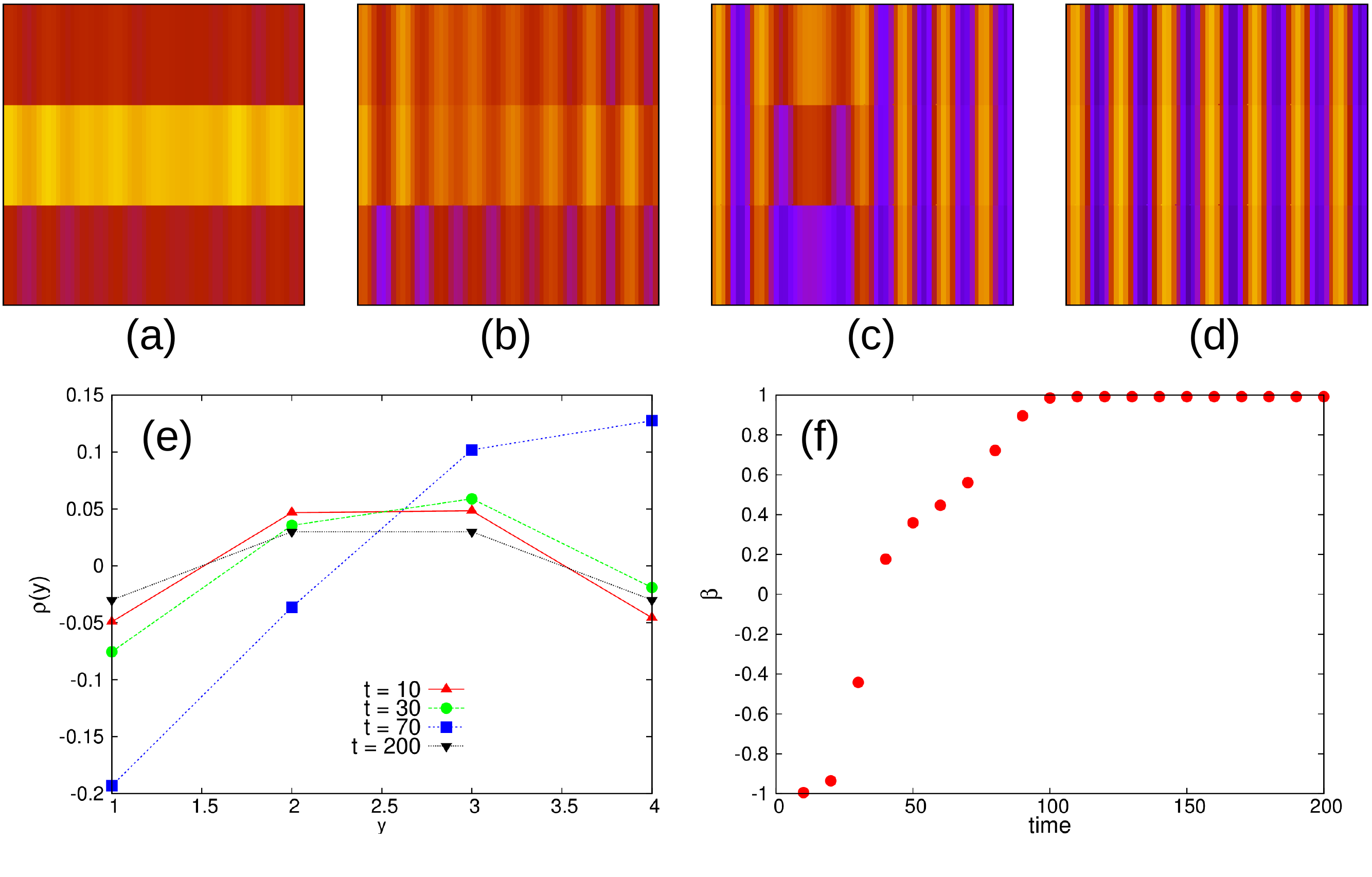}
\caption{\label{fig8}  Microstructure evolution with model parameters $h = 0.1$ and $E = 0.156 $ for film thickness $L_y = 4(0.4L_o)$ at timesteps (a) t = 10, (b) t = 30, (c) t = 70 and (d) t = 200. The microstructure has been drawn in $1:1$ scale for clarity. To relieve the frustration due to confinement, the copolymer arranges in a stable perpendicular configuration even in the absence of electric field. The early time microstructure in (a) is a superposition of parallel and perpendicular lamellae. The final perpendicular state is achieved through a transition state of asymmetric configuration as also seen in the density profile in (e) (t = 70). (f) The degree of alignment parameter corresponds to a perpendicular state.}
\end{figure*}
Next, we decrease the film thickness below the equilibrium lamellar spacing, to $L_y = 4(0.4L_o)$. The results corresponding to model parameters $h = 0.1$ and $E = 0.156$ are shown in Fig. \ref{fig8}. The early stage microstructure corresponding to t = 10 is a superposition of parallel and perpendicular lamellae. With time (t= 30 and 70), the system evolves through a metastable antisymmetric configuration and transforms into a perpendicular state. This observation is also corroborated by the average density profile. The result is consistent with the findings of Walton et al. \cite{walton1994free} who argue in favor of a transient antisymmetric arrangement in symmetric thin films during the formation of vertical configuration.

Even with the smallest electric field strength (used in the course of this study), a perpendicular arrangement is seen to be stable. Even in the absence of the electric field, a stable perpendicular arrangement establishes implying that the geometrical confinement predominates over the electric field. Interestingly, $\beta$ reaches the value of +1 in this case. This observation points towards the following important fact : When the arrangement is guided by substrate confinement, deviation from perfect perpendicular morphology is negligible. However, significant deviation in perpendicularity is observed when the ordering is achieved due to the application of electric field.
\begin{figure*}[ht]
\includegraphics[scale=0.5]{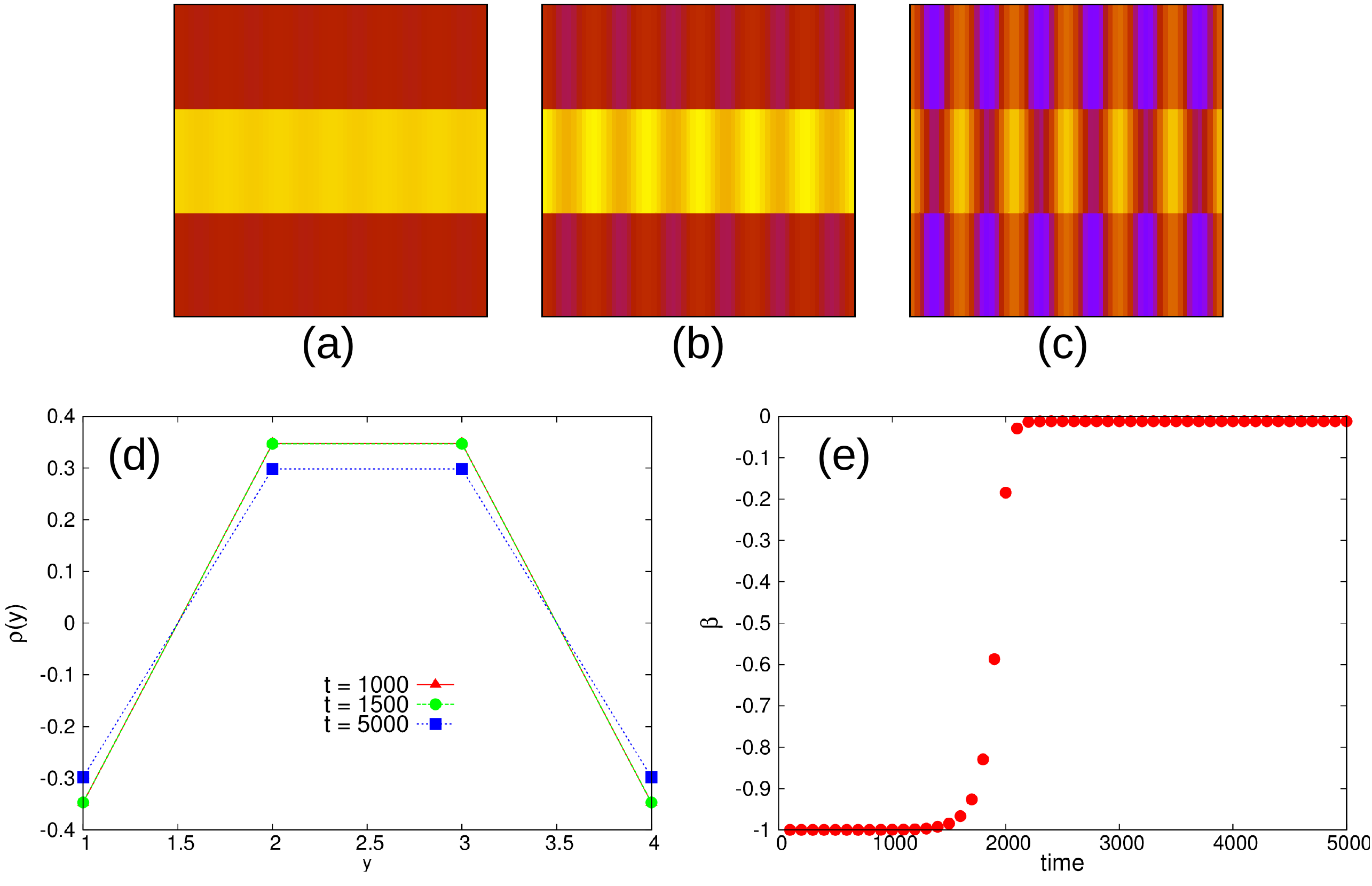}
\caption{\label{fig9}  Microstructure evolution with model parameters $h = 1.0$ and $E = 2.188$ for film thickness $L_y = 4(0.4L_o)$ at timesteps (a) t = 1000, (b) t = 1500 and (c) t = 5000 . The microstructure has been drawn in $1:1$ scale for clarity. The microstructure at final step is a superposition of parallel and perpendicular lamellae. The perpendicular lamellae are thinner at the center. (d) The average density profile is also not flat in the region of $\rho(y) = 0$, implying a significant deviation from perpendicular arrangement. (e) The value of degree of alignment indicates a 50$\%$ aligned structure in the direction of field.}
\end{figure*}
For $h_o \geq 1.0$, parallel arrangement is found to be stable in the absence and at low strength of electric fields. At higher electric field strengths, the microstructure is a superposition of parallel and perpendicular lamellae as can be seen in Fig. \ref{fig9}. Because of the small film thickness, the substrate interaction is predominant and electric field is not able to completely eradicate the previous surface ordering phenomenon. Correspondingly $\beta$ saturates to a value of 0 which is midway between parallel and perpendicular configuration.  
\subsection{Phase diagram}
\begin{figure*}[ht]
\includegraphics[scale=0.8]{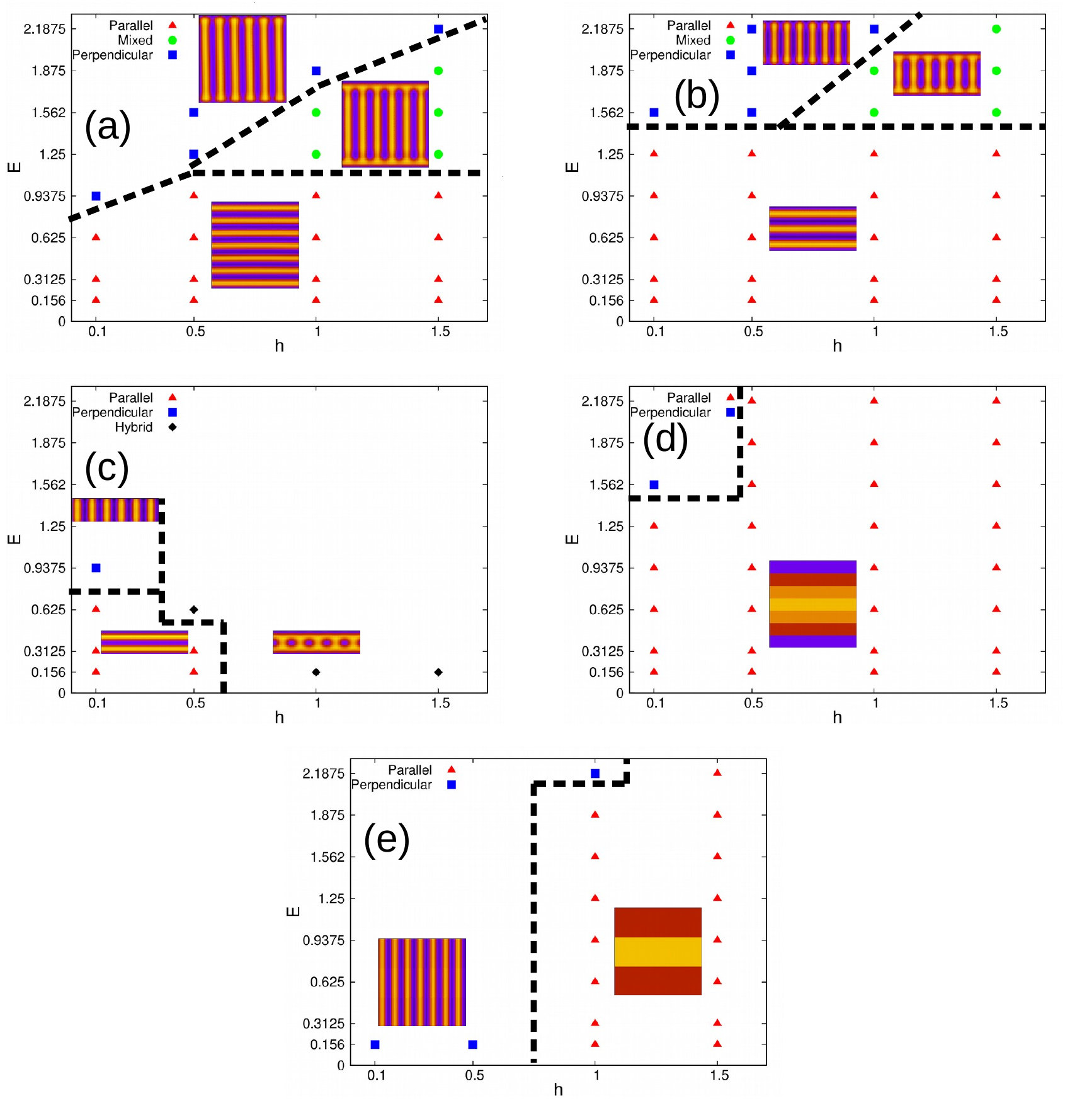}
\caption{\label{fig10} Phase diagram showing the stable arrangement at different magnitudes of electric field $E$ and substrate interaction strength $h$ for different film thicknesses (a) $L_y = 64(6L_o)$, (b) $L_y = 32(3L_o)$, (c) $L_y = 16(1.6L_0)$, (d) $L_y = 8(0.8L_o)$ and (e) $L_y = 4(0.4L_o)$. The microstructures for $L_y = 8$ and $4$ have been drawn in $1:1$ scale. All the above plots are for permittivity values of $\epsilon_A = 3.0$ and $\epsilon_B = 2.0$.}
\end{figure*}

We construct a phase diagram to summarize the influence of electric field $E$, surface interaction strength $h$ and film thickness $L_y$ on the evolving morphologies by classifying them into parallel, perpendicular and mixed category. We designate the morphology mixed only when there exists at least one completely parallel layer ($L = L_o/4$ since the layers closer to the substrate are one-half of the inner layers) . Classification based on such criteria will allow us to compare the resulting phase diagram with the analytical calculations \cite{tsori2002thin, pereira1999diblock}. Any other combination of phases is denoted as hybrid structure. The resulting configuration stability diagram is shown in Fig. \ref{fig10}. The following points can be appreciated,
\begin{enumerate}
\item For a given film thickness, the magnitude of applied electric field to induce a perpendicular arrangement increases with increasing magnitude of substrate interaction strength. Any deviation from this generality occurs only for very thin films e.g. $L_y = 4$ (Fig. \ref{fig10}(e)) and closer to half-integral lamellar thickness i.e $(n+\frac{1}{2})L_o$ (in our case $L_y = 16$, Fig. \ref{fig10}(c)).   

\item At low strength of substrate interaction ($h = 0.1,0.5$) and electric fields, parallel arrangement is found to be stable. However beyond a certain critical value of the electric field e.g. $E = 0.937$ corresponding to $h = 0.1$, the configuration oscillates between perpendicular and parallel configuration (Fig.  \ref{fig10} (a)-(e)). The present findings accentuate the previous analytical results \cite{tsori2002thin}. In general : As $L_y$ decreases, the effect of substrate interaction becomes more prominent. Therefore a higher magnitude of electric field is required to induce a transition from parallel to perpendicular configuration. However, for $L_y$ incommensurable with the bulk lamellar spacing (say $(n + \frac{1}{2})L_o$) i.e. halfway between integral lamellar spacings, the free energy of the parallel configuration is maximum and hence a lower electric field can induce a perpendicular transition.

\item For unstrained films e.g. $L_y = 64, 32$, the critical electric field $E_c$ required for a parallel to perpendicular transition scales as $L_y^{-1/2}$ \cite{tsori2002thin}. Though the numerical calculations are carried out at discrete values of electric field, we can still verify this behavior: $E_c$ for h = 0.1 and $L_y = 64$ lies between $0.625$ and $0.9375$ (Fig. \ref{fig10} (a)). Taking $E_c = 0.9$ for $L_y = 64$, $E_c$ for $L_y = 32$ can be calculated to be $1.27$ which lies between the values of $1.25$ and $1.562$ in Fig. \ref{fig10}(b). Similar trend can be verified for $h = 0.5$ in Fig. \ref{fig10}(a) and (b).

\item For higher substrate interaction strengths, i.e $h = 1.0, 1.5$, the range of stability of parallel configuration increases with decreasing film thickness (Fig. \ref{fig10}(a), (b), (d), (e)). For $L_y < L_o$, the parallel configuration is more stable.

\item Mixed morphologies are stable only for thicker films ($L_y \geq 3L_o$) and for higher substrate interaction strength, $h \geq 1$ (Fig. \ref{fig10}(a), (b)). Interestingly, the electric field required for the transition of parallel to mixed morphology, i.e. the first critical field is independent (or at most weakly dependent due to the discretized nature of the phase diagram) of substrate interaction strength (Fig. \ref{fig10}(a) and (b)). However, the second critical field i.e. the field required to convert mixed to perpendicular morphology is dependent strongly upon the substrate interaction strength (approximately linearly). 

\item Next, we consider the variation of the critical fields with film thickness for unstrained films ($L_y = 64,32$) for fixed substrate interaction strengths of $h = 1.0,1.5$ (Fig. \ref{fig10} (a) and (b)). Clearly, both the critical fields depend upon the film thickness. The dependence of the first critical field, though, is stronger than the second critical field.

\item Interesting morphologies arise at $L_y \sim 1.5L_o$ where the film thickness is incommensurable with the lamella period. At low substrate affinities a usual parallel to perpendicular lamellae transition is observed but at higher affinities a wide range of hybrid structures results. The exact nature, though, is not clear in the present study. 

\end{enumerate}
\begin{figure}[ht]
\includegraphics[scale=0.5]{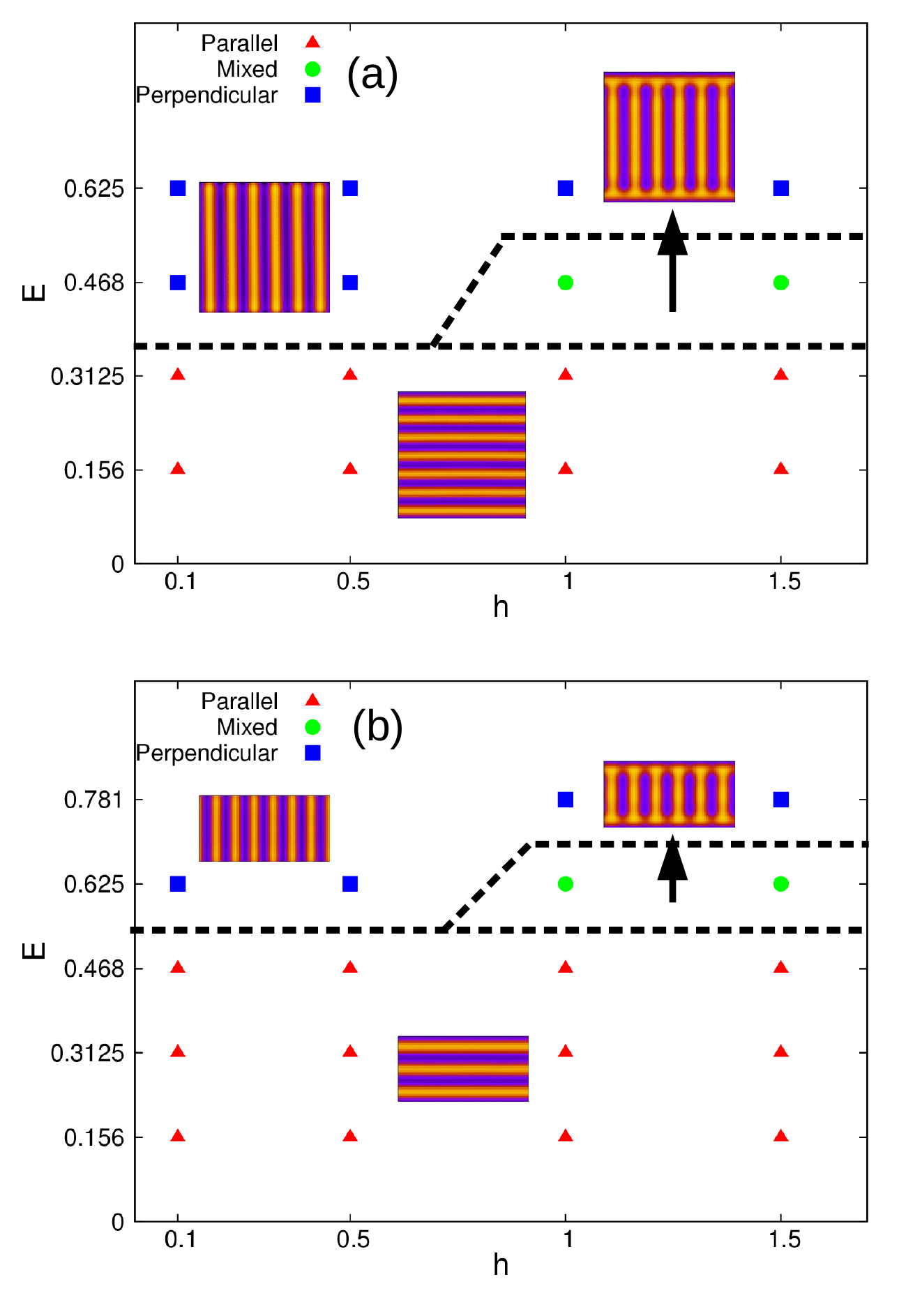}
\caption{\label{fig11} Phase diagram showing the stable arrangement at different magnitudes of electric field $E$ and substrate interaction strength $h$ for film thickness (a) $L_y = 64(6L_o)$, (b) $L_y = 32(3L_o)$ for permittivity values of $\epsilon_A = 6.0$ and $\epsilon_B = 2.5$.}
\end{figure}
\subsection{Role of dielectric contrast}
The effect of increasing the dielectric contrast i.e. $\epsilon_A - \epsilon_B$ on the final configuration, will now be discussed. Increasing the contrast between the two blocks implies that the material is more responsive to an applied electric field and, as a consequence, if $(\epsilon_A- \epsilon_B)$ is large, the resulting phase diagram is governed by the relative mismatch. In the studies so far, we consider $\Delta \epsilon = 1$ ($\epsilon_A = 3$ and $\epsilon_B = 2$), which are close to the values reported by Amundson et al. \cite{amundson1994alignment} ($\epsilon_A = 3.8$ and $\epsilon_B = 2.5$) for PS-PMMA copolymer. If we now increase the permittivity difference $\Delta \epsilon = 3.5$ by selecting $\epsilon_A = 6.0$ and $\epsilon_B = 2.5$ as reported in Ref. \cite{lyakhova2006kinetic, tsori2002thin, matsen2006undulation}, for the same copolymer system, an increase in electrostatic free energy contribution is expected for the same magnitude of applied field. 

We restrict the discussion to unstrained films ($L_y = 64,32$). The resulting phase diagram is presented in Fig. \ref{fig11}  and the corresponding values of the critical electric fields are drastically lowered. According to the analytical calculations \cite{tsori2002thin, pereira1999diblock}, this decrease is proportional to $\frac{\sqrt{\epsilon_A + \epsilon_B}}{\epsilon_A - \epsilon_B}$. Considering the critical electric field to be $E_c = 0.9$ for $\Delta \epsilon = 1$ and $L_y = 64$ and $h = 0.1$ (Fig. \ref{fig10}(a)), the critical field on increasing the dielectric contrast to $\Delta \epsilon = 3.5$ according to above equality (on holding $L_y$ and $h$ constant) yields a critical value of around $E_c = 0.3$ which complies well with Fig. \ref{fig11}(a). A similar behaviour is retrieved for $L_y = 32$ as well.

The nature of the phase diagram changes dramatically and the region of mixed morphology in the phase diagram is diminished. With enhanced dielectric contrast, the dependence of the critical fields on the substrate interaction strength for a given film thickness becomes rather weak. This is contrary to the behavior at low dielectric contrast where the second critical field (mixed to perpendicular) displayed a strong dependency on substrate interaction strength. However, the dependency of the critical fields on film thickness for a given substrate interaction strengths is similar to that at low dielectric contrast i.e. both fields depend on the film thickness with the dependency of the second critical field being higher than the first.

\begin{figure*}[ht]
\includegraphics[scale=0.4]{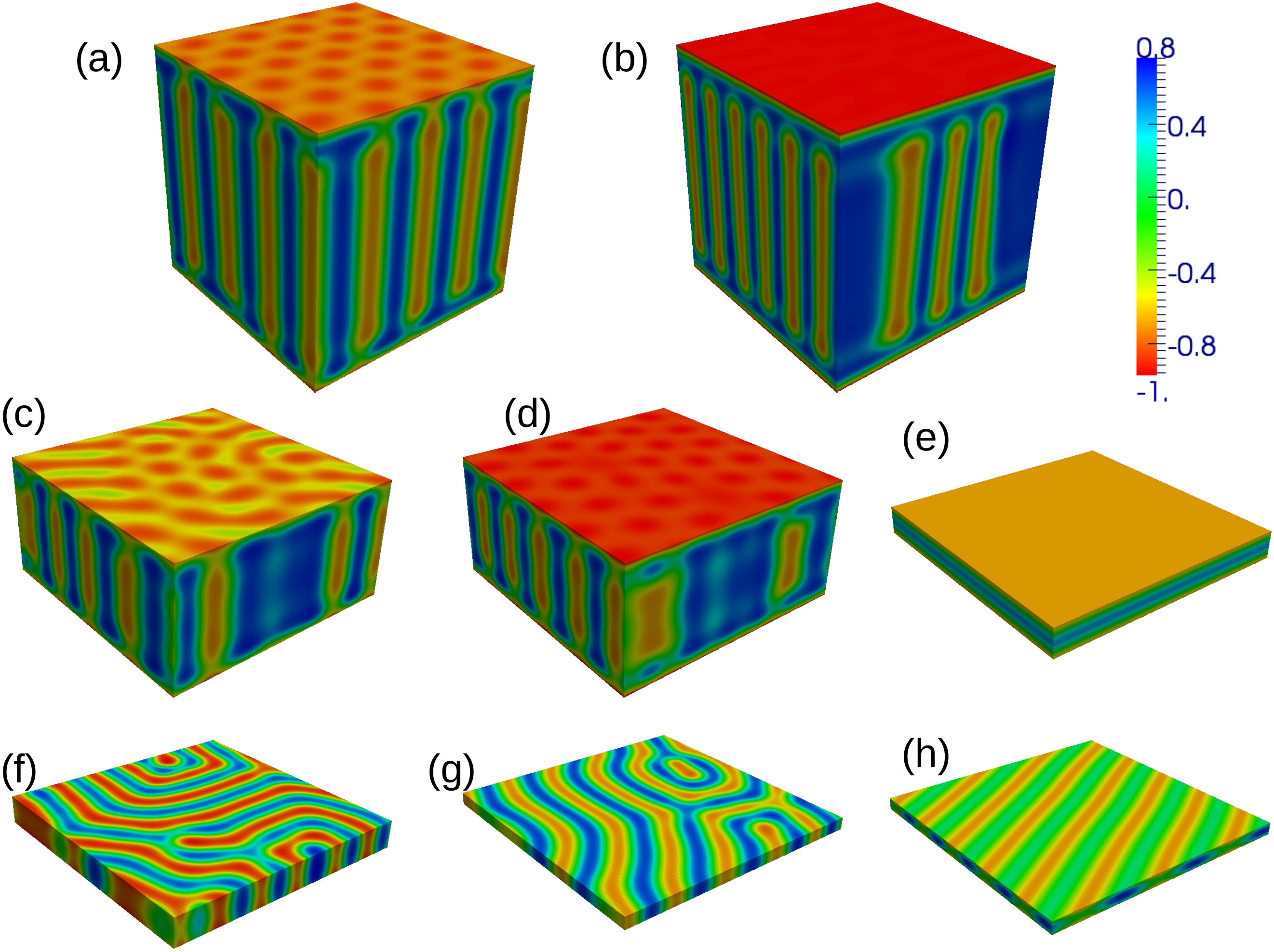}
\caption{\label{fig12} Microstructural patterns at (a) $L_y = 64(6L_o)$,$h = 0.5$, $E = 1.25$, (b) $L_y = 64(6L_o)$,$h = 1.0$, $E = 1.25$, (c) $L_y = 32(3L_o)$,$h = 0.5$, $E = 1.562$, (d) $L_y = 32(3L_o)$,$h = 1.0$, $E = 1.562$, (e) $L_y = 8(0.8L_o)$,$h = 0.1$, $E = 1.25$, (f) $L_y = 8(0.8L_o)$,$h = 0.1$, $E = 1.562$, (g) $L_y = 4(0.4L_o)$,$h = 0.1$, $E = 0.156$, (h) $L_y = 4(0.4L_o)$,$h = 1.0$, $E = 2.1875$. A comparison with the 2D morphologies in the phase diagram presented in Fig.\ref{fig10} suggests that the microstructure fall into the same adopted morphology classification in 3D as well.}
\end{figure*}
\begin{figure*}[ht]
\includegraphics[scale=0.5]{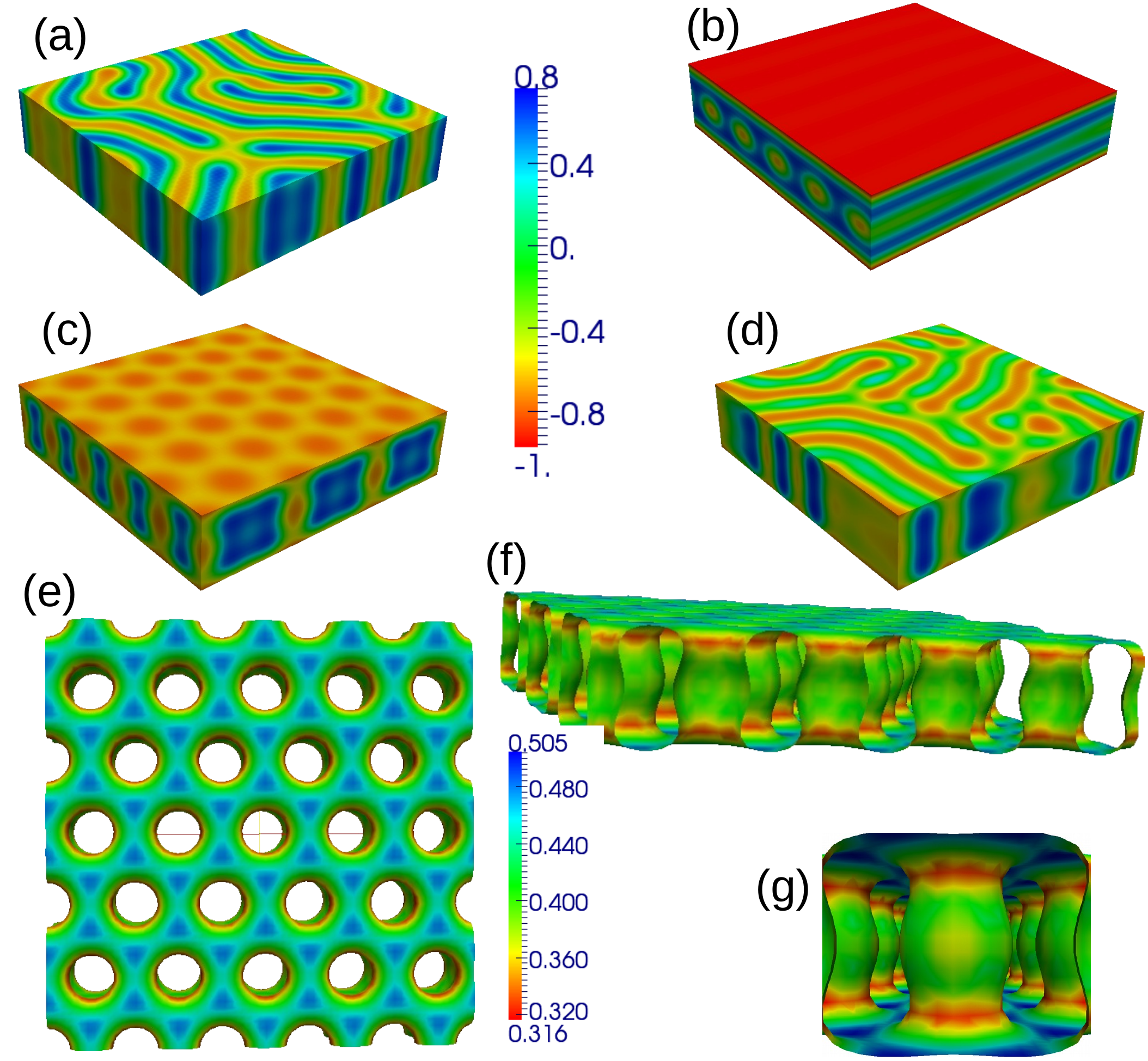}
\caption{\label{fig13} Microstructural patterns for $L_y = 16(1.5L_o)$ at substrate affinity and electric field strength of (a) $h = 0.1$, $E = 0.9375$, (b) $h = 1.0$, $E = 0.156$, (c) $h = 0.5$, $E = 1.25$, (d) $h = 0.5$, $E = 2.1875$, (e) top view of the contour of the order parameter gradient of (c), (f) sideview of (e) and (g) a zoomed view of (f). The coexistence of parallel wetting layers and an inner parallel layer of cylinders can be seen in (b).The presence of electric field induces the formation of perpendicular cylinder in (c). However at higher field strengths, the system reverts to a perpendicular lamellae (d).}
\end{figure*}
\section{Discussions and Conclusions}\label{sec4}
We conclude the paper with a  critical assessment of the results from the present study to the experiments, analytical and SCFT calculations. The most relevant experimental study is due to Xu et al. \cite{xu2003interfacial, xu2004electric}. They studied thin films of varying thickness ($4L_o, 10L_o, 20L_o, 100L_o$) \cite{xu2004electric}. For film thicknesses less than $10L_o$, substrate interaction was found to be dominant resulting in parallel arrangement, even on application of electric field of $40V/\mu m$ . The substrate induced ordering length in their study was about $5L_o$ and the segmental interaction or the segregation was $\chi N = 26$. The segregation in the present work is relatively weaker than their experiments ($\chi N = 18$) and ordering length is around $8L_o$ (with lowest substrate interaction strength). The critical field for lowest substrate interaction from our study is of the order of $76V/\mu m$ and $128V/\mu m$ for $L_y = 6L_o$ and $3L_o$ respectively for a dielectric contrast of $\Delta \epsilon = 1$. The second value is actually quiet high and may well exceed the dielectric breakdown of the material and in such cases only parallel ordering would be exhibited. We remark that in recent experiments, electric field of the order of $120V/\mu m$ have been applied \cite{crossland2010control}. However, if we examine the critical field for enhanced dielectric contrast of $\Delta \epsilon = 3.5$ (the same as in experiment), the values are around $40V/\mu m$ and $52V/\mu m$ for $L_y = 6L_o$ and $3L_o$ respectively, which are well within experimental range. Given a higher segregation in experiments, we can expect the critical fields to be higher than that in the present study. Infact using SCFT Matsen \cite{matsen2006undulation}  calculated the critical field to be around $57V/\mu m$ for the same experimental conditions of Xu et al. ($\Delta \epsilon = 3.5$ and $\chi N = 26$) for $L_y = 10L_o$, though substrate interaction was not explicitly considered. Moreover if we compare the critical field for $\chi N = 18$ from their work, the critical field would turn about to be roughly $58V/\mu m$ for dielectric contrast of $\Delta \epsilon = 3.5$ and $L_y = 6L_o$(Fig 5(a) in Ref. \cite{matsen2006undulation}).

We additionally compare our findings to that of Lin et al. \cite{lin2006self} who studied sphere to cylinder transition. Using SCFT calculations, a complete phase diagram was calculated.  Their segregation also corresponds to $\chi N = 18$. The present work is thus complementary to their study. For weak substrate interaction, the critical field (maximum value) calculated by them is around $32V/\mu m$ and $45V/\mu m$ for $L_y = 6L_o$ and $3L_o$ respectively (Fig. 7 in ref. \cite{lin2006self}). Considering that sphere to cylinder transition generally takes place at field strength lower than parallel to perpendicular lamella transition, our values of $40V/\mu m$ and $52V/\mu m$ are quite in agreement with SCFT calculations. A similar comparison can also be made at higher substrate strengths.

The calculated phase diagrams are similar in spirit to that by Lyakhova et al. \cite{lyakhova2006kinetic}. In both studies, the phase diagram is obtained from dynamic microstructure evolution rather than static calculations \cite{tsori2002thin, lin2006self}. The authors investigated parallel to perpendicular transition of lamellar morphology using dynamic SCFT coupled to perturbed solution of Maxwell equation for thin films of $L_y = 4L_o$ and segregation of $\chi N = 16$. Mixed morphologies were however not observed in that study. In the present study, the system was allowed to evolve from a disordered state under combined electric and substrate field, whilst in the study of Lyakhova et al., electric field was applied to well developed microstructures. Possible difference can arise because of the initial level of ordering. Our results are qualitatively similar to the results of Lin et al. \cite{lin2006self} who observed the presence of mixed phases in film thickness as low as $3L_o$ in cylinder forming systems at a similar segregation.

We next compare our phase diagram to the analytical calculation of Tsori et al. \cite{tsori2002thin}. The authors computed the phase diagrams both in weak and strong segregation regime. The results presented correspond to an intermediate regime. Though our results are closer to WSL, the phase diagram presented in Fig . \ref{fig10} is similar to the analytical calculation of Tsori et. al in SSL (Fig. 8 and 9 in ref. \cite{tsori2002thin}). We remark that the two phase diagrams correspond to two different segregation regimes and are based on different assumptions. Phase diagram calculated by Tsori et al.  corresponds to SSL and is based on the assumption of finite surface ordering length and high dielectric contrast ($\Delta \epsilon = 3.5$ as compared to our $\Delta \epsilon = 1$). In WSL, they assumed the ordering length to be greater than the film thickness and mixed morphology was not considered. In the present study, mixed morphology is observed inspite of the ordering length being lower than the film thickness. We believe that the consideration of finite surface ordering length in context of WSL (and in the occurence of mixed morphology in particular) might have been an over assumption. Moreover, only a single intermediate phase i.e. a mixed lamellae morphology was considered. The results of the present study, however, indicates the presence of other intermediate or hybrid structures, thus altering the phase diagram significantly from analytical theories.

We conclude by briefly discussing the influence of wall interaction characteristics on the equilibrium morphologies. In the present work, we have restricted the study to symmetric substrate interaction. However, in principle two additional cases are possible. The substrate can be (i) antisymmetric i.e. both walls attract different monomers with same strength or (ii) asymmetric i.e. both walls have preference towards same or different monomers, but possess different interaction strength. We believe that the consideration of antisymmetric case may not lead to any new geometries other than the ones reported here. Only the region of their respective occurrence in the phase diagram might change, given that film thickness corresponding to integral number of lamellae spacing would then be the frustrated state and half-integral, the natural state. On the contrary, asymmetric interaction can potentially engender an additional type of mixed morphology, where the system adopts a parallel configuration on one side (where interaction strength is comparatively higher) and perpendicular configuration on the other (where interaction strength is weaker). However, it remains to be seen if the competing electric field can stabilize other morphologies, for instance, cylindrical that has been reported in the present study. 

To summarize, we have studied the morphology evolution of a symmetric diblock copolymer under competing substrate interaction and electric field using a coupled Ohta-Kawasaki functional and Maxwell equation. By solving the full Maxwell equation, we do not assume weak dielectric inhomogeneity, making the model equally applicable irrespective of segregation. A good agreement with the analytical and SCFT calculations, amply demonstrates the predictive capability of the proposed model. A distinct advantage of coarse graining is the accessability to large scale simulation, especially in three dimension. Moreover, in 3D the nature of mixed/hybrid morphologies are well defined and such a simple classification (parallel, perpendicular and mixed lamellae) may not be suffice. Our 2D results does point out that in the incommensurate films in the regime $L_y<2L_o$, this interplay of substrate, confinement and electric field leads to rich hybrid structures and even the occurence of a parallel lamellae to perpendicular cylinder transition. Infact in recent experiment on gyroid forming copolymers by Crossland et al. \cite{crossland2010control} a large number of coexisting morphologies were observed at low dielectric contrast. We present some preliminary 3D results in Fig. \ref{fig12} and \ref{fig13} which corroborates the findings of our present 2D study. The exact nature of these hybrid structures and their transition in electric field will be communicated shortly.


%
%

%


\begin{acknowledgments}
The work has been supported financially by the ministry of the state Baden-Wuerttemberg
(Mittelbau programme).
\end{acknowledgments}
\end{document}